\documentclass[aps,prb,twocolumn,showpacs,floatfix,superscriptaddress]{revtex4-1}
\usepackage{dcolumn}
\usepackage{bm}
\usepackage{soul}
\usepackage{amsmath,amssymb,graphicx}
\usepackage[colorlinks=true,citecolor=blue,urlcolor=blue,linkcolor=blue]{hyperref}
\usepackage[caption=false]{subfig}

\newcommand{\eV}{\,\mathrm{eV}}
\newcommand{\nB}{n_\mathrm{B}}
\newcommand{\nD}{n_\mathrm{D}}
\newcommand{\kF}{k_\mathrm{F}}
\newcommand{\meV}{\,\mathrm{meV}}

\newcommand{\kk}{\mathbf{k}}

\newcommand{\qq}{\mathbf{q}}

\newcommand{\trel}{t_\mathrm{rel}}
\newcommand{\tave}{t_\mathrm{ave}}

\newcommand{\td}{t_\mathrm{delay}}
\newcommand{\fs}{\,\mathrm{fs}}
\newcommand{\resigma}{\mathrm{Re}\Sigma^{\mathrm{R}}}
\newcommand{\imsigma}{\mathrm{Im}\Sigma^{\mathrm{R}}}

\newcommand{\imdr}{\mathrm{Im}\ D^R}

\newcommand{\imgr}{\mathrm{Im}\ G^R}

\newcommand{\CC}{\mathcal{C}}
\newcommand{\TT}{\mathcal{T}}
\newcommand{\FF}{\mathrm{F}}
\DeclareGraphicsExtensions{.png,.jpg,.pdf}
\begin{document}

\title{Nonequilibrium Electron Dynamics In Pump-Probe Spectroscopy: Role Of Excited Phonon Populations}

\author{O.~Abdurazakov}
\email{oabdura@ncsu.edu}
\affiliation{Department of Physics, North Carolina State University, Raleigh, NC 27695}
\author{D.~Nevola}
\affiliation{Department of Physics, North Carolina State University, Raleigh, NC 27695}
\author{A.~Rustagi}
\affiliation{Department of Physics, North Carolina State University, Raleigh, NC 27695}
\author{J.~K.~Freericks}
\affiliation{Department of Physics, Georgetown University, Washington, DC 20057, USA}
\author{D.~B.~Dougherty}
\affiliation{Department of Physics, North Carolina State University, Raleigh, NC 27695}
\author{A.~F.~Kemper}
\email{akemper@ncsu.edu}
\affiliation{Department of Physics, North Carolina State University, Raleigh, NC 27695}

\date{\today}
%%%%%%%%%%%%%%%%%%%%%%%%%%%%%%%%%%%%%%%%%%%%%%%%

\begin{abstract}
We study the role of excited phonon populations in the relaxation rates of nonequilibrium electrons using a nonequilibrium Green's function formalism. The transient modifications in the phononic properties are accounted for by self-consistently solving the Dyson equation for the electron and phonon Green's functions. The pump induced changes manifest in both the electronic and phononic spectral functions. We find that the excited phonon populations suppress the decay rates of nonequilibrium electrons due to enhanced phonon absorption. The increased phonon occupation also sets the nonequilibrium decay rates and the equilibrium scattering rates apart. The decay rates are found to be time-dependent, and this is illustrated in the experimentally observed population decay of photoexcited $\mathrm{Bi}_{1.5}\mathrm{Sb}_{0.5} \mathrm{Te}_{1.7}\mathrm{Se}_{1.3}$.     
\end{abstract}
%%%%%%%%%%%%%%%%%%%%%%%%%%%%%%%%%%%%%%%%%%%%%%%%

\maketitle

%%%%%%%%%%%%%%%%%%%%%%%%%%%%%%%%%%%%%%%%%%%%%%%%
\section{Introduction}

Driving matter far from equilibrium is a new frontier in the control of quantum materials. Prominent recent discoveries including Floquet insulators \cite{xia} and Time Crystals \cite{wilczek,rovny} herald important opportunities to create new phenomena and materials properties by purposefully driving materials away from their well-known equilibrium states.  A decisive factor in the properties and stability of driven, nonequilibrium matter is the relaxation of excited degrees of freedom.  We need to quantify the time scales of thermalization and relaxation, and to identify the rate limiting steps in thermalization and relaxation processes as targets for characterization and control.  This focus on driven matter necessitates new experimental and theoretical tools capable of addressing the time-dependent phenomena.  On the theoretical side, the fundamental challenge is that time translation symmetry is broken out of thermal equilibrium and full quantum dynamics calculations need to be performed for which methodology is still in development \cite{KemperAssump}.  On the experimental side, the challenge is that the most relevant relaxation processes in quantum matter occur on ultrafast time scales in the range of femtoseconds to picoseconds \cite{bove}.  

Of the quintessential techniques to study matter out of equilibrium, time-resolved photoelectron spectroscopy is rapidly becoming an established probe of quantum materials. With
increasing resolution and breadth of application, time- and angle-resolved photoemission spectroscopy (tr-ARPES)
is used as a tool to study materials such as high-T$_c$ cuprates,\cite{cuprates,cuprates2,cuprates3, ShenARPES} graphene,\cite{graphene,bigraphene} and other 2D materials.\cite{MoS2,trARPES2D}
In these experiments, one of the common measurements is that of population dynamics.\cite{giannetti,Bauer, MoS2dynamics} 
A laser pump excites the electrons within the material, which absorb energy and occupy states above the Fermi level that were unoccupied in equilibrium.  With time, the electrons relax back to a new final state, which may be the same as the pre-pump equilibrium state or a modified one. Recent work examining the return of excited systems to equilibrium in tr-ARPES experiments\cite{SentefPRX,KemperPRB,Entropy} demonstrated that a rate-limiting step in the energy transfer from nonequilibrium electrons to the phonon bath, may set the rate of relaxation even in the presence of other interactions such as impurity and Coulomb scattering.\cite{RameauNatComDissip,KemperAssump}
This is also reflected in the fact that quasiparticle lifetimes (as measured, e.g. through an ARPES linewidth) and population
decay rates (as measured using tr-ARPES) are often inequivalent.\cite{YangPRL}

However, in these earlier works the phonon bath was assumed to have infinite heat capacity i.e. the phonon properties(e.g., frequency, linewidth, occupation) remained unchanged. While this is a reasonable approximation when the amount of absorbed energy is small,
in principle the phonon bath may absorb energy, and may transfer energy back to the electrons. The recent simultaneous measurement of the electron and lattice dynamics in optically excited systems has also demonstrated that the mutual energy transfer between the electrons and the phonons determine the relaxation dynamics.\cite{Konstantinova} 
Furthermore, considering the modification of phonon properties out of their equilibrium state is critical, especially when the
phonons are driven directly. This possibility has been of interest within the field of light-induced states of matter,\cite{Mitrano,fausti,rini}
and more specifically in reports of light-enhanced electron-phonon coupling upon phonon driving\cite{Gierz}.
Within the context of tr-ARPES, one may expect electron population scattering rates to decrease as the phonon occupation increases (zero at low temperatures) because this causes increased phonon absorption and thus may slow down the decay. 

The theoretical study of relaxation dynamics has been of interest since the advent of ultrafast spectroscopy. Subsequent developments have been heavily influenced by the so-called two-temperature model (TTM) suggested by Allen\cite{Allen} which helped to understand the relaxation dynamics in simple metals after an ultrafast excitation. The model assumes that the electrons and the lattice are independently thermalized and are characterized by electron and lattice temperatures; the difference in their respective temperatures drives the dynamics. As the time-resolution of the probe pulses improved, thermalization and the relaxation processes could no longer be decoupled, and they often overlap in time\cite{Nonthermal}. To account for this, the research focus shifted from solving the semi-classical Boltzmann equation numerically\cite{BTE, Rethfeld, Maldonado} and analytically\cite{Kabanov} to solving microscopic models using advanced numerical techniques such as time-dependent exact diagonlization\cite{Golez},  nonequilibrium Greens' function\cite{SentefPRX, KemperPRB}, and time-dependent DMFT\cite{NEDMFT, Werner} methods.  Using the latter method, which is applicable for the systems of higher dimensions/coordination number, Murakami \textit{et. al.} have highlighted the importance of the phonon dynamics during the relaxation toward equilibrium.\cite{Murakami} When the phonon dynamics are taken into account, they find a qualitatively different relaxation dynamics of quenched populations where there is a crossover from electron to phonon dominated relaxation at different coupling regimes, explained in terms of the different dependence of the electron and phonon self-energies on the electron-phonon coupling strength. However, the recent work\cite{KemperAssump} demonstrates numerous cases where the self-energy can no longer fully dictate the population relaxation dynamics.   

\begin{figure}
	\includegraphics[width=0.99\columnwidth]{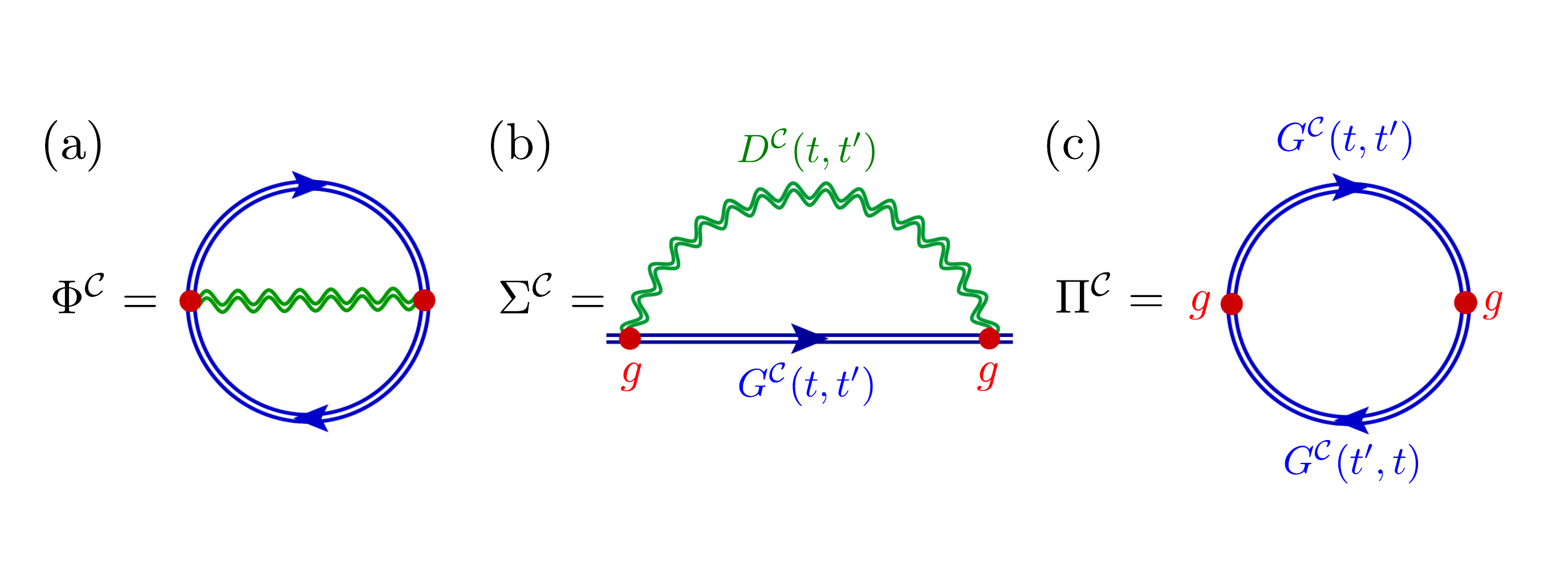}
	\caption{Pictorial depiction of $\mathrm{(a)}$ the functional, the contour $\mathrm{(b)}$ electron and $\mathrm{(c)}$ phonon self-energies. A double-line denotes the self-consistently renormalized Green's function.}
	\label{fig:diagrams}
\end{figure}

In this work, we investigate the effect of self-consistently renormalizing the phonon properties on the relaxation dynamics by considering a fully conserving approximation. As illustrated in Fig.~\ref{fig:diagrams}, we use functionally-derived self-energies for the electrons and phonons, consistent with a conserving approximation.  This allows the mutual energy transfer between electrons and phonons.
We solve the equations of motion for the system using a non-equilibrium Green's function (NEGF) method,
as detailed in Sect.~\ref{sec:model}. 

Comparison of the nonequilibrium electron properties with those in equilibrium helps us to clarify the role of the enhanced (excited) phonons in (out of) equilibrium. Our results reveal several novel aspects of the population dynamics that are otherwise absent in cases where the phonon properties stays fixed.  Due to the increase in phonon population, the relaxation decay rate is slowed down, and may even change sign within a phonon energy of the Fermi level (termed the ``phonon window'') as the increasing phonon population pushes quasiparticles from below the Fermi energy to above.  We investigate these effects as a function of excitation density and probe delay time. We also show that the pump field can effectively modify the signatures of interactions in both electron and phonon spectra.

The paper is organized as follows.  In Sec.~\ref{sec:model} we discuss our model and method.  In Sec.~\ref{subsec:equilib} we present the equilibrium electronic quantities followed by the analysis of the simulated tr-ARPES spectra and the nonequilibrium sum rules in Sec.~\ref{subsec:arpes}.  Then in Sec.~\ref{subsec:decay_rates} we obtain the decay rates from
the tr-ARPES spectra, and discuss their binding energy- and time-dependence. In support of the latter, we show experimentally measured time-dependent decay rates of the excited surface states of $\mathrm{Bi}_{1.5}\mathrm{Sb}_{0.5} \mathrm{Te}_{1.7}\mathrm{Se}_{1.3}$. In Sec.~\ref{subsec:phonons} we illustrate the effect on the phononic
spectra. We conclude in Sec.~\ref{sec:conclusion}.

%%%%%%%%%%%%%%%%%%%%%%%%%%%%%%%%%%%%%%%%%%%%%%%%

\section{Model and Method}
\label{sec:model}

We study the dynamics of the electrons residing in a 2D tight-binding band linearly coupled to a bath of optical phonons of a frequency $\Omega_0$. The system is described by the Holstein Hamiltonian\cite{holsteinI, holsteinII}
\begin{align}
        \mathcal{H} =&  \sum_{\kk\sigma} \epsilon_{\kk}c_{\kk,\sigma}^{\dagger}
        c_{\kk,\sigma} + \sum_{\qq} \Omega_0 b_{\qq}^{\dagger}b_{\qq}\nonumber\\
 +&\textit{g} \sum_{\kk\qq \sigma} c_{\kk+\qq, \sigma}^{\dagger}c_{\kk, \sigma}(b_{\qq}+b_{-\qq}^{\dagger}),
\end{align}
where $c_{\kk_\sigma}(b_{\qq})$ and $c_{\kk,\sigma}^{\dagger}(b_{\qq}^{\dagger})$ are the electron (phonon) annihilation and creation operators at the state $\kk (\qq)$ and a spin $\sigma$. The electron-phonon interaction vertex $g$ is assumed to be momentum-independent.  
The tight-binding band dispersion is given by 
\begin{align}
        \epsilon_{\textbf{k} } = -2V_\mathrm{nn}(\cos k_x + \cos k_y) + 4 V_\mathrm{nnn} \cos k_x\cos k_y - \mu.
\end{align}
The electrons can hop between the (next) nearest neighboring sites with the $(V_\mathrm{nnn}) V_\mathrm{nn}$ amplitude, and the chemical potential $\mu$ determines the band filling.  
 
The electronic system is driven from the initial equilibrium by a pump field described by the vector potential $\mathbf{A}(t)$. We include the pump field via the Peierls' substitution $\kk\rightarrow \kk - \mathbf{A}(t)$,\cite{Peierls} which is spatially uniform. This method of including the external field allows to have both the electron-phonon scattering processes and the ultrafast optical excitation to take place at the same time as opposed to interaction quench methods\cite{Murakami} or the numerical integration of Boltzmann transport equations.\cite{BTE} In our choice of units where $k_B = c = \hbar = e  = 1$, and working within the Hamiltonian gauge, the electric field of the pump is given by $\mathbf{E}(t) = -\partial_t \mathbf{A}(t)$ with zero scalar potential.

We solve the equation of motion for the system using a nonequilibrium Green's function method whose main object is the double-time Green's function living on the Keldysh contour $\mathcal{C}$. For an electron and phonon, the Green's functions are defined by\cite{Mahan} 
\begin{align}
G_{\kk}^{\CC}(t,t') &= -i\langle \TT_\CC c_{\kk}(t)c_{\kk}^{\dagger}(t') \rangle,\\
D_\qq^{\CC}(t,t') &= -i\langle \TT_\CC X_\qq(t)X_\qq^{\dagger}(t') \rangle,
\end{align} 
respectively, where $\TT_\CC$ is the time-ordering operator on the contour $\mathcal{C}$.  The phonon displacement and creation(annihilation) operators are related via $X_\qq = b_\qq + b_{-\qq}^{\dagger}$. Since we work with Einstein phonon modes, the phonon Green's function is manifestly local. The thermal average is taken over the initial equilibrium distribution at temperature $T$. 
The Green's function $G_{\kk}^{\CC}(t,t')$ evolves on the Keldysh contour according to the Dyson integro-differential equation
\begin{align}
        [i\partial_t - \epsilon_{\textbf k}(t)]G_{\textbf k}^{\mathcal C}(t,t')= \delta^{\mathcal C}(t,t') + \displaystyle \int_{\mathcal C}{\textit dz}\Sigma^{\mathcal C}(t, \textit z) G_{\textbf k}^{\mathcal C}(\textit z,t').\label{eq:KB}
\end{align}
The electron self-energy $\Sigma^{\mathcal C}$ accounts for the effect of electron-phonon interactions.  

The phonon Green's function is obtained by solving its Dyson equation
\begin{align}
D^{\mathcal C}(t, t') &= D_{0}^{\mathcal C}(t,t ')\nonumber\\
&+\iint_{\CC} dt_1dt_2D_0^{\mathcal C}(t,t_1)\Pi^{\mathcal C}(t_1, t_2)D^{\mathcal C}(t_2,t').\label{eq:dyson}
\end{align}
Here, $\Pi^{\mathcal C}(t_1, t_2)$ is the phonon self-energy.
The noninteracting phonon Green's function $D_{0}^{\mathcal C}(t,t ')$ is given by\cite{Mahan}
\begin{align}
D_{0}^{\mathcal C}(t,t ') = -i[&(n_B(\Omega_0)+1-\theta_\CC(t,t'))e^{i\Omega_0 (t-t')}\nonumber\\
                          +    &(n_B(\Omega_0)+\theta_\CC(t,t'))e^{-i\Omega_0 (t-t')}].
\end{align}
Here, $n_B(\Omega_0) = (e^{\Omega_0/T}-1)^{-1}$ is the Bose distribution function, and $\theta_\CC$ is the Heaviside step function on the Keldysh contour. The details of the method used to solve the equations of motion are outlined elsewhere\cite{algorithm}.

In this work, we consider two cases. First, we work with phonons where the phonon Green's function is kept fixed, i.e., $D^\CC\equiv D^\CC_0$.
 In this approximation, phonons serve as an effective heat bath with infinite heat capacity which will be denoted as ``Infinite Bath'' for brevity. In the second case, we solve the Dyson equation for the phonon Green's function self-consistently as we move forward in time which takes into account the transient modifications of the phonon properties as they interact with electrons. Thus phonons in this case resemble a heat bath with a finite heat capacity which will be denoted as ``Finite Bath''. To solve Eq.~\ref{eq:KB} and Eq.~\ref{eq:dyson}, we need to chose an approximation scheme for the electron and phonon self-energies. In the case of the Finite Bath, we use a conserving approximation where the particle number, momentum, and the total energy of the system are conserved. For an approximation to be conserving, the self-energies $\Sigma$ and $\Pi$ have to be functional derivatives of a Littinger-Ward functional $\Phi[G,D]$\cite{Baym62}. The functional and the self-energy diagrams are depicted in Fig.~\ref{fig:diagrams}. The Hartree term in the electron-self energy is absorbed into the chemical potential as it is momentum independent and instantaneous. The corresponding expression for the self-energy diagrams are
\begin{align}
\Sigma^{\CC}(t,t') &= ig^2D^{\CC}(t,t')G_{\mathrm{loc}}^{\CC}(t,t'),\\
\Pi^{\CC}(t,t) &= -ig^2G_{\mathrm{loc}}^{\CC}(t,t')G_{\mathrm{loc}}^{\CC}(t',t)\label{eq:phonon_self},
\end{align}
where the local electron Green's function $G_{\mathrm{loc}}^{\CC} = N_{\kk}^{-1}\sum_{\kk}G_{\kk}^{\CC}$.  Here, we ignore the momentum dependence of the phonon self-energy by summing over all electronic momenta in the first Brillouin zone. Often, the optical phonon bands are relatively dispersionless so this is a good approximation. When the acoustical branches are also considered, the momentum dependence becomes relevant. However, for the relatively short time dynamics we are interested in, the dominant relaxation channel is provided by the high energy optical phonons.

Once we have chosen the appropriate self-energy approximation,  we can solve Eqs. \ref{eq:KB} and \ref{eq:dyson}. By applying the Langreth rules,\cite{stefanucci_book} one can separate the contour equation into the pieces residing in the different parts of the Keldysh contour, i.e. Matsubara, real-time, and mixed pieces,\cite{Mahan} and standard numerical integration techniques can be employed\cite{algorithm}. 
%%%%%%%%%%%%%%%%%%%%%%%%%%%%%%%%%%%%%%%%%%%%%%%%%
\begin{figure*}
	%\fbox{
	\includegraphics[clip=true, trim=0 0 0 0, width=0.90\textwidth]{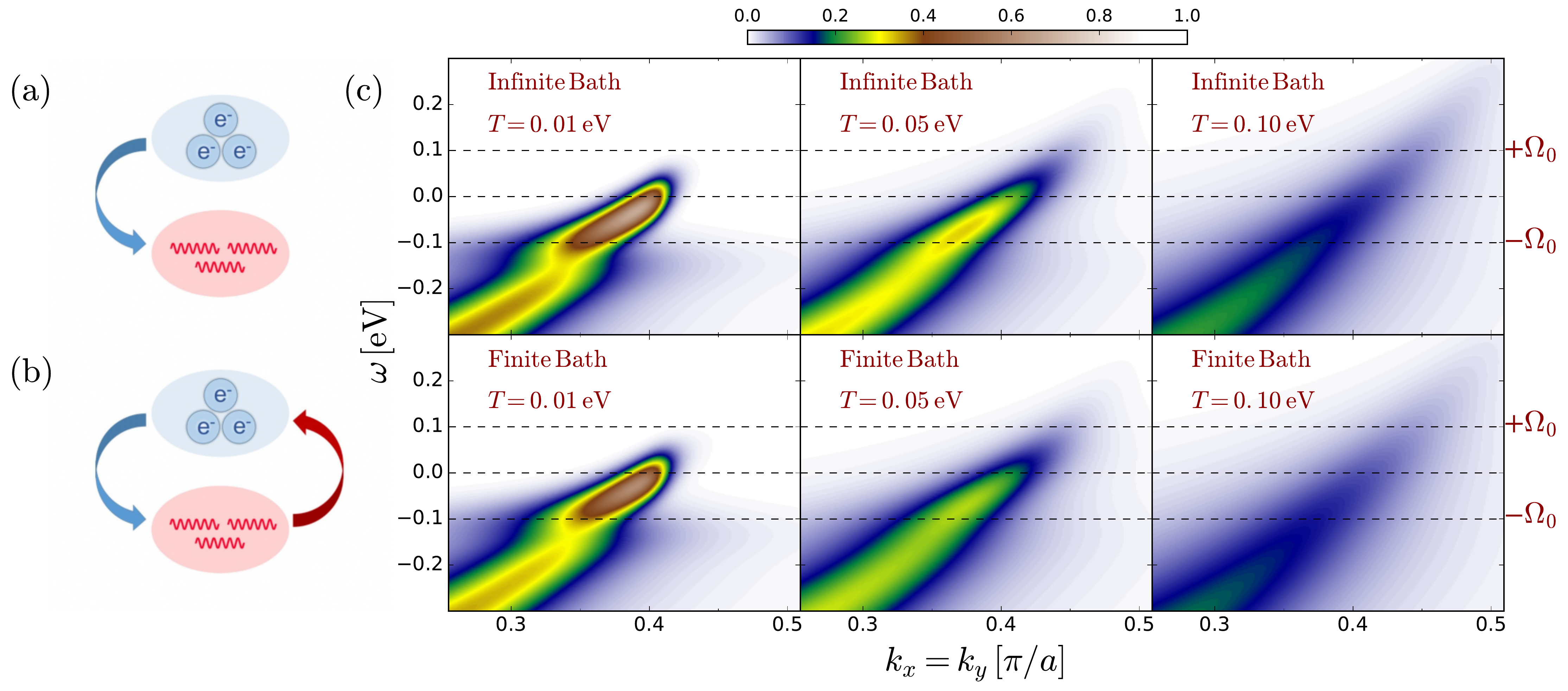}
	%}
	\caption{\textbf{Equilibrium}: Modifications of the equilibrium electronic properties due to temperature. $\mathrm{a,b)}$ Schematic of electrons coupled to the Infinite Bath $\mathrm{(a)}$ and the Finite Bath (b) of phonons. The arrow indicates the possible directions for energy transfer. $\mathrm{c)}$ ARPES spectra along the zone diagonal at different temperatures for both infinite and finite phonon baths. The horizontal dashed lines indicate the bare phonon frequency. The equilibrium temperature increases from left to right.}
	\label{fig:arpes_eq}
\end{figure*}
%%%%%%%%%%%%%%%%%%%%%%%%%%%%%%%%%%%%%%%%%%%%

The pump field centered at $t_0$ and directed along the zone diagonal in the 2D Brillouin zone has a profile given by
\begin{align}
\mathbf{A}(t) = {\mathrm F}\mathrm{sin}(\omega_\mathrm{p}t)e^{-(t-t_0)^2/2\sigma_p^2}(\mathbf{\hat e}_x + \mathbf{\hat e}_y),
\end{align}
where $\FF$, $\omega_{\mathrm p}$, and $\sigma_{\mathrm p}$ are the pump fluence, oscillation frequency, and temporal width, respectively. Here $\mathbf{\hat e}$ is a unit vector along the respective direction in reciprocal space.
We measure the tr-ARPES intensity of electrons using a probe pulse of the Gaussian time profile of width $\sigma_{\mathrm pr}$ 
via\cite{TheoryARPES}
\begin{align}
I(\kk,\omega,t_0) = &\frac{1}{\sqrt{2\pi}\sigma_{\mathrm pr}}\mathrm{Im} \int dtdt'e^{i\omega(t-t')}G_{\kk}^<(t,t')\nonumber\\
& \times e^{-(t-t_0)^2/2\sigma_{\mathrm pr}^2}e^{-(t'-t_0)^2/2\sigma_{\mathrm pr}^2}.
\end{align}
As the measured tr-ARPES intensity must be gauge-invariant, the momentum shift induced by the pump field is corrected by a time-dependent shift.\cite{gauge_correction}

To obtain the energy- and time-dependence of various observables, we perform a Wigner transformation of the time coordinates $\{t,t'\}\rightarrow \{\tave = (t+t')/2,\trel=t-t'\}$. Consequently, the energy dependence is acquired by Fourier transforming with respect to the relative time ($\trel$). The average time ($\tave$) is the measure of the probe delay time with respect to the center of the pump pulse ($\td = \tave - t_0$). 
 
In equilibrium, the coupling strength between electrons and phonons can be quantified by a dimensionless parameter $\lambda = -\partial \mathrm{Re}[\Sigma^\mathrm{R}(\omega)]/\partial\omega \vert_{\omega=0}$. 
For a broad band ($W>>\Omega_0$),  its value depends on the coupling vertex, the density of states at the Fermi level, and the phonon frequency as $\lambda = 2 g^2 N(0)/\Omega_0$.
    
We choose the band parameters to be $V_\mathrm{nn} = 0.25\,\eV$, $V_\mathrm{nnn} = 0.075\,\eV$ , and $\mu = -0.255 \eV$. 
The phonon frequency $\Omega_0= 0.1 \eV$ and the coupling vertex $ g = \sqrt{0.02} \eV^{-1}$ 
are chosen so that we stay in the weak coupling limit ($\lambda < 1$). 
To resolve the spectral features in tr-ARPES, we set the initial temperature of the system to $T = 0.01\,\eV$ or about $116 \mathrm{K}$.

The pump field has a width of $\sigma_{\mathrm p}  \approx 6.6 \fs$ and frequency $\omega_{\mathrm p} = 0.5 \eV$. To resolve the pump induced changes in the electronic as well as phononic spectra, we use a pump fluence of $\mathrm{F}=1.0/\mathrm{a}_0$ ($\mathrm{a}_0$ is the lattice constant), and to compare the nonequilibrium decay rates to their equilibrium counterpart, we employ weaker pump fluences in the range of $0.50/\mathrm{a}_0 \leq \FF \leq 1.00/\mathrm{a}_0$. Henceforth, we set $\mathrm{a}_0 =1$. For better energy resolution of the spectral features, the probe width is taken to be $\sigma_{\mathrm pr} \approx 16.5\fs$.

%%%%%%%%%%%%%%%%%%%%%%%%%%%%%%%%%%%%%%%%%%%%%%%%

\section{Results}

To set the stage for the nonequilibrium results, we present the properties of electrons at thermal equilibrium with the phonon bath at various system temperatures. Then, we focus on nonequilibrium to identify the similarities and the key differences between the spectra of electrons coupled to the Infinite Bath and the Finite Bath both in and out of equilibrium. 

\subsection{Equilibrium}\label{subsec:equilib}

\begin{figure}
	\includegraphics[width=0.99\columnwidth]{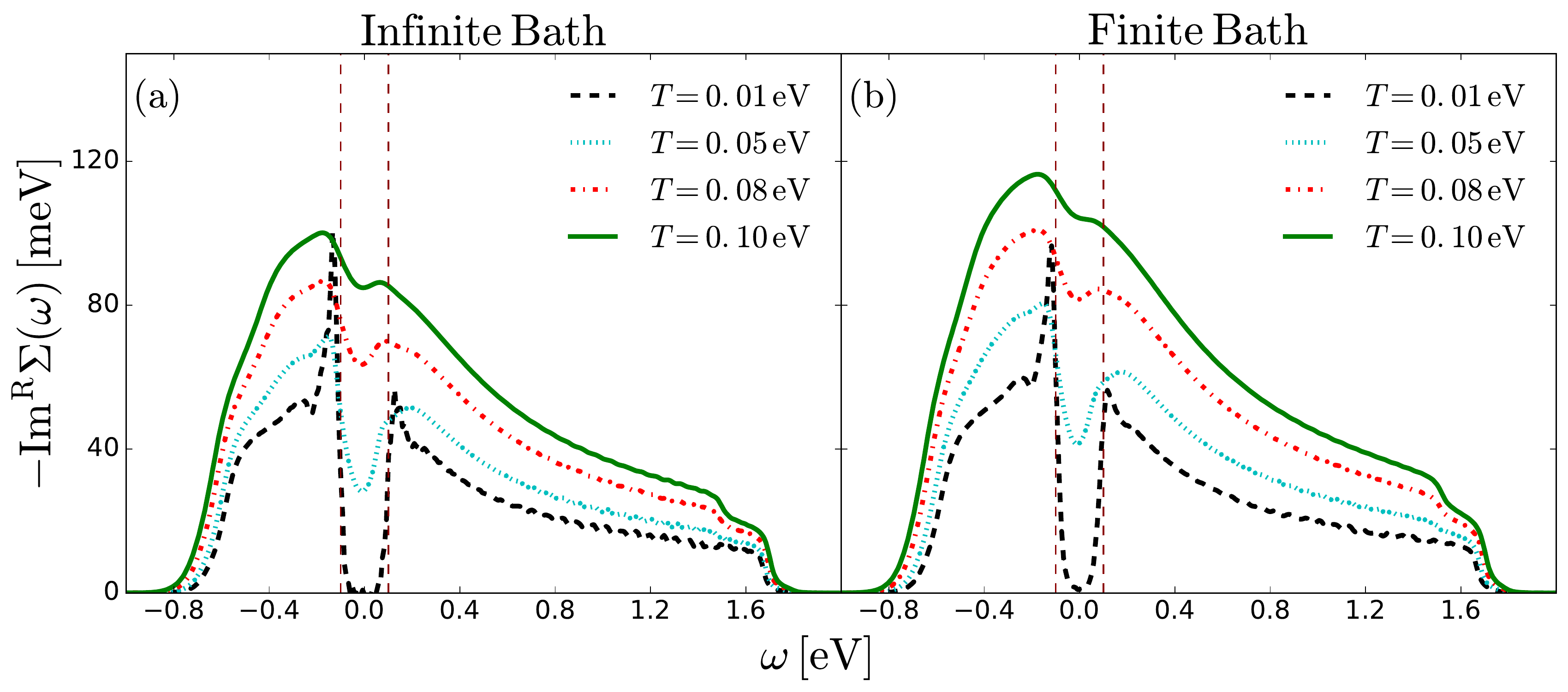}
	\caption{Changes in the imaginary part of the electron self-energy due to increasing temperature for (a) the Infinite Bath and (b) the Finite Bath. The phonon-window (demarcated by the vertical dashed lines) is filled up with increased spectral weight as the temperatures is increased.}
	\label{fig:imsigma_equilib}
\end{figure}

The equilibrium ARPES directly accesses the electronic states with a given energy resolution. To aid the contrast between the spectra of electrons coupled to the Infinite Bath and to the Finite Bath we include the cartoons representing the two cases. The cartoons depict the systems of electrons coupled to the Infinite Bath in Fig.~\ref{fig:arpes_eq}$\mathrm{(a)}$ and the Finite Bath in Fig.~\ref{fig:arpes_eq}$\mathrm{(b)}$. The arrows indicate the possible directions of energy flow between the electronic and phononic subsystems.
 
Fig.~\ref{fig:arpes_eq}(c) is the comparison of the ARPES spectra of the electrons coupled to the Finite Bath and the Infinite Bath in the vicinity of the characteristic kink at different temperatures. The kink, traditionally, is regarded as a measure of the strength of the electron-boson interaction because its size and shape are determined by the electron self-energy.\cite{ashcroft} The magnitude of the spectral-weight at various quasi-particle energies and momenta is reflected in the intensity of the spectra. We mark the bare phonon frequency by horizontal dashed lines.

\begin{figure}
	\includegraphics[width=0.99\columnwidth]{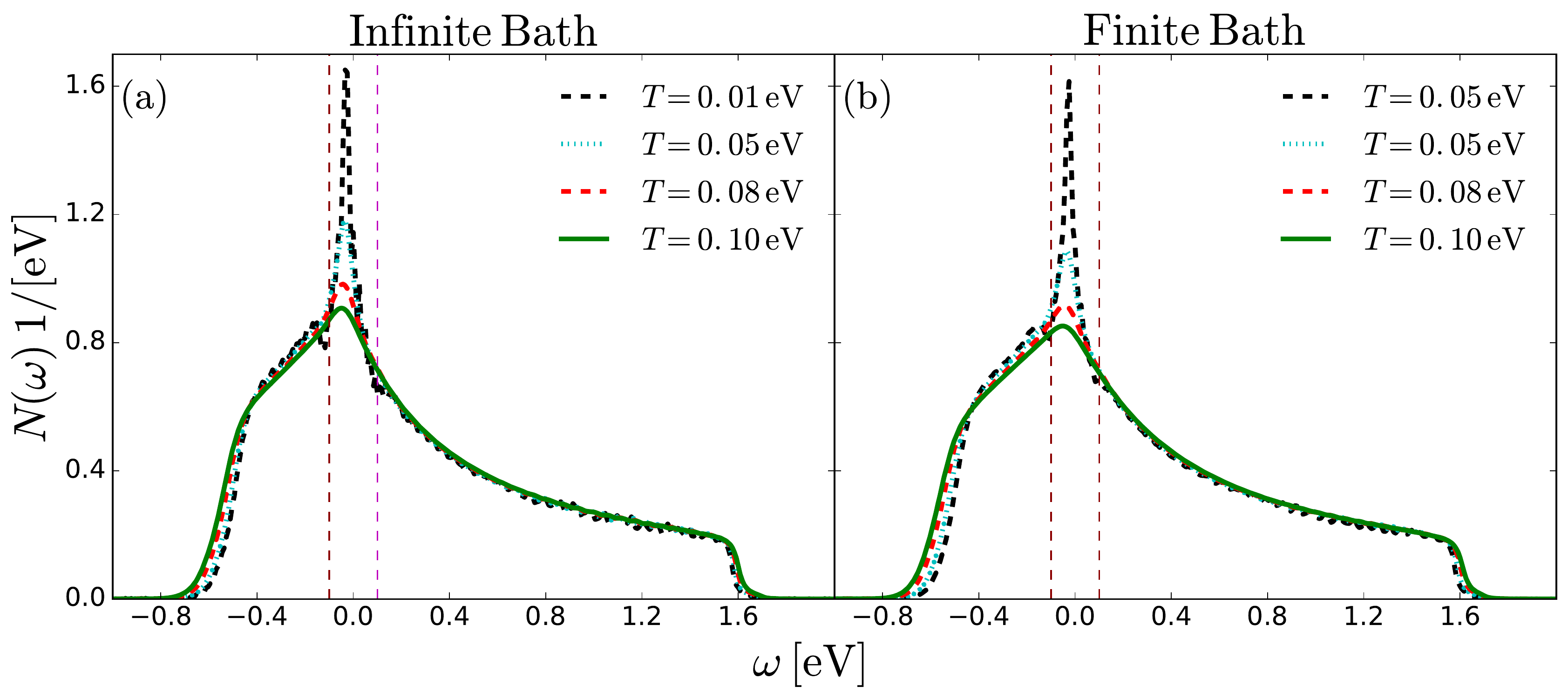}
	\caption{Changes in the electron density of states due to temperature for (a) the Infinite Bath and (b) the Finite Bath. Increased temperature causes noticeable decrease for the energies $|\omega|<\Omega$.}
	\label{fig:dos_equilib}
\end{figure}

The equilibrium spectra at $T=0.01\eV$ for both cases exhibit well defined kinks around the Fermi level corresponding to an interacting electron-phonon system. At this temperature, the phonon population is vanishingly small. The spectra, in the case of the Finite Bath, display a slightly shifted kink toward the Fermi level indicating a renormalized phonon frequency due to interactions with electrons. The subsequent panels show that increasing the system temperature causes the spectra to acquire an apparently weakened kink and diffuse spectra. These two properties are connected to the real and the imaginary part of the electron self-energy.\cite{Mahan} The diffuseness of the spectra or the magnitude of the spectral linewidth reflects the changes in $\imsigma$ as shown in Fig.~\ref{fig:imsigma_equilib} for various temperatures for the Infinite Bath as well as the Finite Bath. This quantity is responsible for the scattering rates of the quasiparticles. Since the electrons are coupled to the optical phonons of frequency $\Omega$, there is a phase space constraint for an electron having a lower energy than $\Omega$ causing low scattering rates in this so-called phonon-window. This is demonstrated by the black dashed curve in the figure for $T=0.01\eV$ where the phonon-window effect is most prominent due to reduced thermal smearing. As we increase the system temperature the rates inside the phonon-window increase as more phonons become available for scattering. By comparing the spectra of the Finite Bath to those of the Infinite Bath, we conclude that treating the phonon properties self-consistently affects the single-particle spectra noticeably, but not significantly. We see the expected broadening of the spectra and the loss of the kink feature.

The equilibrium spectral-weight measured by the angle integrated photoemission spectroscopy is the product of the electron distribution function $f(\omega)$ and the electronic density of states  $N(\omega)$ obtained from the retarded component of the electron Green's function whose poles and width are effected by $\resigma$ and $\imsigma$ as $N(\omega) = -\frac{1}{\pi} \imgr_\mathrm{loc}(\omega)$ . This quantity is shown in Fig.~\ref{fig:dos_equilib} for various temperatures. Changing the temperature modifies the density of states inside the phonon-window and at the band edges though the total integrated area enclosed by $N(\omega)$ does not change and is equal to unity.\cite{SumRules} The sharp feature in the vicinity of the Fermi level (the van Hove singularity) at low temperatures is flattened out as the system temperature is increased. This indicates a decrease in the number of the available states in the aforementioned region. 

\subsection{Nonequilibrium}

\subsubsection{tr-ARPES}\label{subsec:arpes}

\begin{figure*}
	%\fbox{
	\includegraphics[clip=true, trim=0 0 0 0, width=0.90\textwidth]{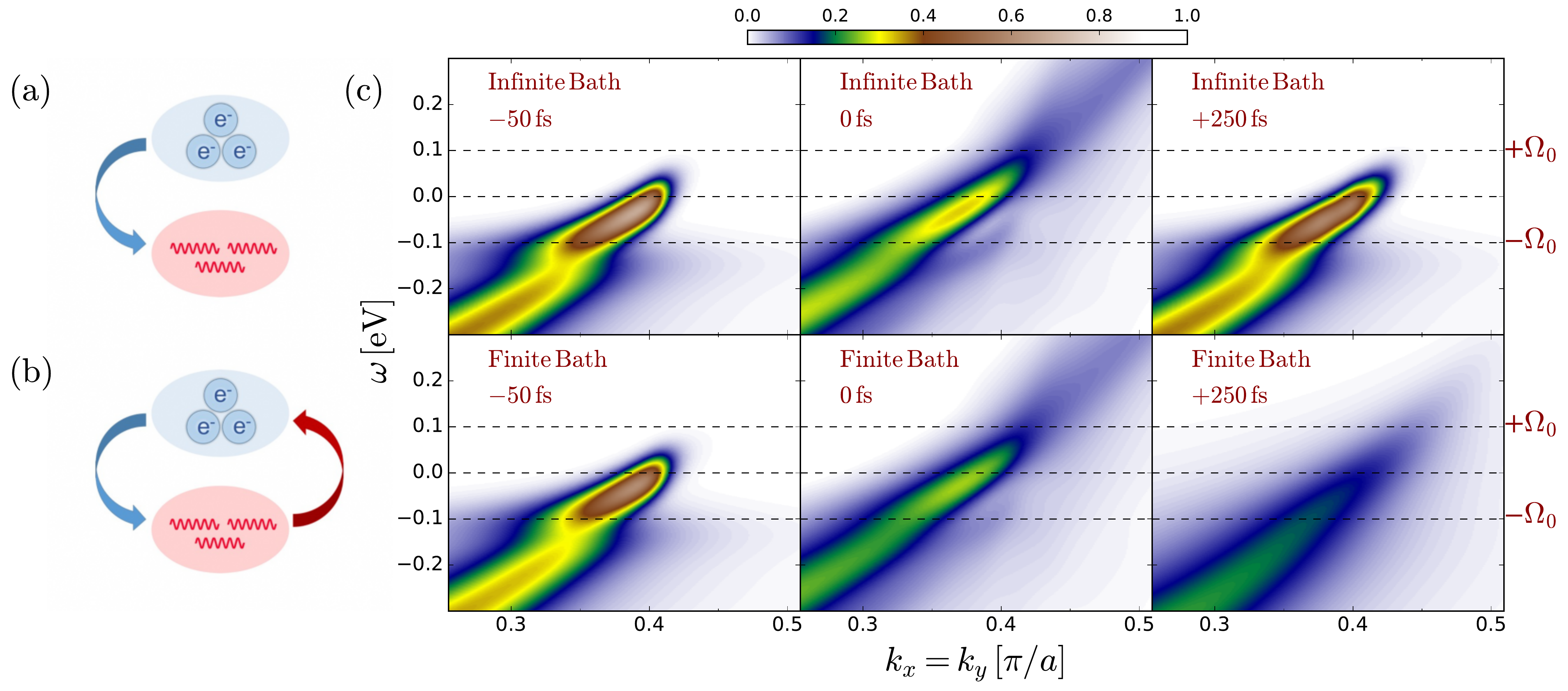}
	%}
	\caption{\textbf{Nonequilibrium}: Temporal evolution of electronic properties. $\mathrm{a,b)}$ Schematic of electrons coupled to the Infinite Bath $\mathrm{(a)}$ and the Finite Bath (b) of phonons. The arrow indicates the possible directions for energy transfer. $\mathrm{c)}$ tr-ARPES spectra along the zone diagonal before(equilibrium), during, and long time after the pump for both infinite and finite phonon baths. The horizontal dashed lines indicate the bare phonon frequency. The initial equilibrium temperature is set to $T = 0.01\eV$, and the pump fluence used to drive the system $\mathrm{F}=1.0/\mathrm{a}_0$.}
	\label{fig:arpes}
\end{figure*}

\begin{figure*}
	%\fbox{
	\includegraphics[clip=true, trim=0 0 0 0, width=0.90\textwidth]{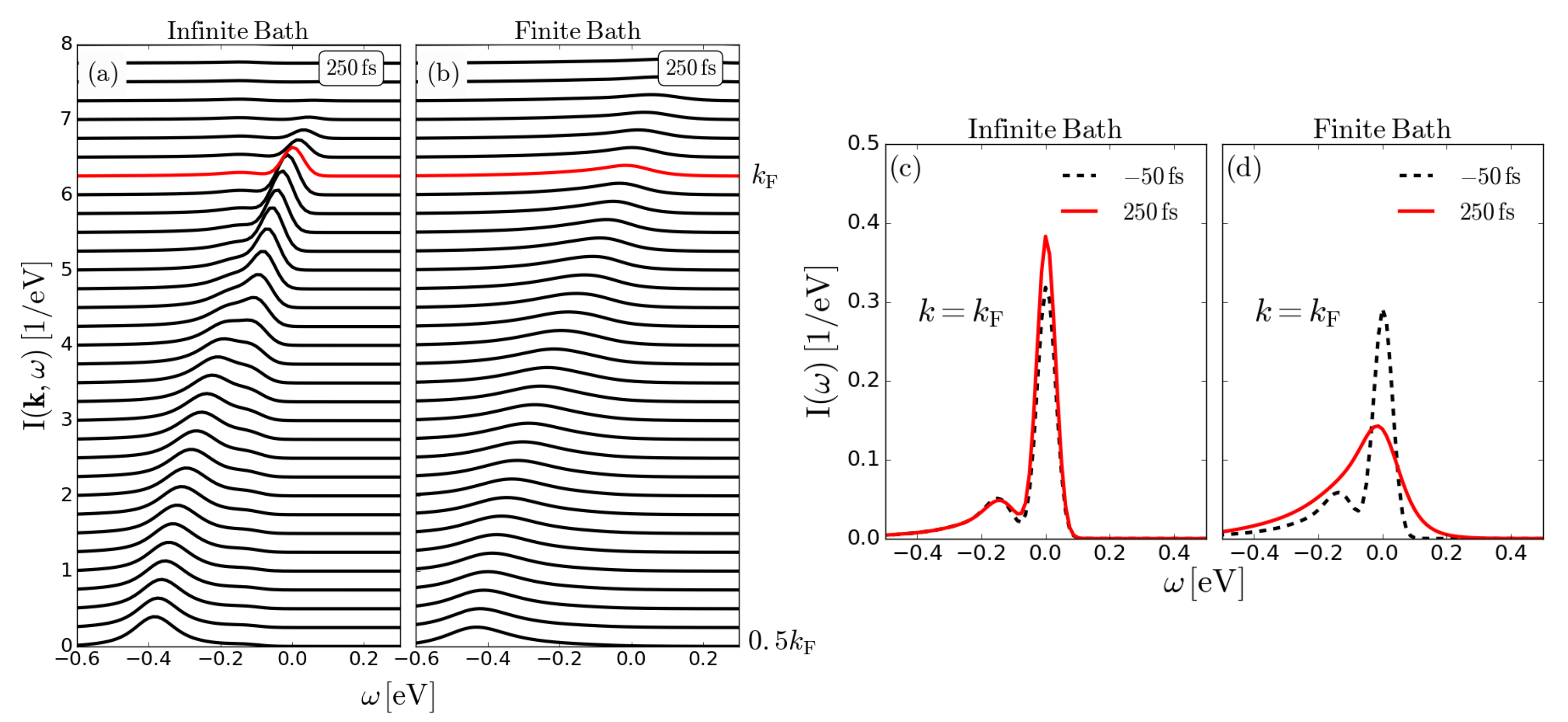}
	%}
	\caption{Comparison of the energy distribution curves (EDC) of (a) the Infinite Bath to those of (b) the Finite Bath. The EDCs at various quasiparticle momenta are put side-by-side to highlight the changes in the electronic spectra due to the excited phonons. The EDCs at the Fermi momentum long after the pump pulse compared to those in equilibrium for (c) the Infinite Bath as well as (d) the Finite Bath.}
	\label{fig:waterfall}
\end{figure*}

In the nonequilibrium realm, we note that the phonon population (set by the equilibrium lattice temperature) stays fixed in the Infinite Bath, whereas it can be excited in the Finite Bath. The latter is possible because of the energy supplied by the pump is conserved and dynamically transferred between the electrons and the Finite Bath resulting in the increase of the phonon population. In Fig.~\ref{fig:arpes}, we compare the tr-ARPES of the electrons interacting with the Finite Bath to those interacting with the Infinite Bath in equilibrium, just after, and a long time after the excitation. We are interested in how the modified phononic properties affect the single particle dynamics as well as the population dynamics of electrons.  
Initially, the coupled system of electrons and phonons are in thermal equilibrium. Before excitation, the electrons mostly occupy states below the Fermi level except for a small amount of spectral-weight 
because of the nonzero initial temperature. Once the pump pulse excites a significant portion of the spectral 
weight above the Fermi level($\td=0\fs$), the spectra of electrons coupled to the Infinite Bath show a disappearance of the kink resulting from a rearrangement of spectral-weight by the pump field in agreement with the previous results.\cite{KemperPRB} Long after the excitation ($\td = 250\fs$), the system returns to its pre-pump equilibrium form.

On the other hand, the electrons interacting with the Finite Bath respond to the ultrafast excitation differently. In addition to the disappearance of the kink, the excited electrons also manifest a more diffuse spectra. The long time form of the spectra also differ starkly from those of the Infinite Bath. It does not return to its initial equilibrium form, rather it settles to a different final state where the kink is no longer visible and the linewidth is broadened.

To illustrate these long-time differences, we plot the energy distribution curves (EDC) at various momenta in Fig.~\ref{fig:waterfall}(a/b) for the Infinite/Finite Bath. We observe clear differences. The ones for the Finite Bath acquire a more smeared form at all momenta. This is associated with the increased phonon population during the electron relaxation because the spectral linewidth is directly related to the electron self-energy (see Fig.~\ref{fig:imsigma_equilib}), which becomes more smeared as the phonon temperature is increased. In the right figure, we compare EDC curves for the initial and final states for both cases. In the case of the Infinite Bath, the EDC at $\kF$ essentially restores to its pre-pump form, while the kink fades away in the case of the Finite Bath.

Although the notion of temperature is not strictly defined for the nonequilibrium electrons, to understand the features in tr-ARPES we refer back to the equilibrium spectra in Fig.~\ref{fig:arpes_eq} at different temperatures. At delay times long enough for electrons and the phonons to reach thermal equilibrium, the single particle spectra takes a form analogous to the equilibrium spectra at an elevated temperature for the Finite Bath and the initial temperature for the Infinite Bath. However, in the case of the Finite Bath, there is no way to know the final system temperature \textit{a priori}. We note that the transient states can never be obtained from equilibrium calculations even with the knowledge of the respective electron and phonon temperatures (assuming that the population are thermal which does not strictly hold)\cite{Entropy,KemperAssump} because those two temperatures differ at a given delay time.
\begin{figure}
	\includegraphics[width=0.80\columnwidth]{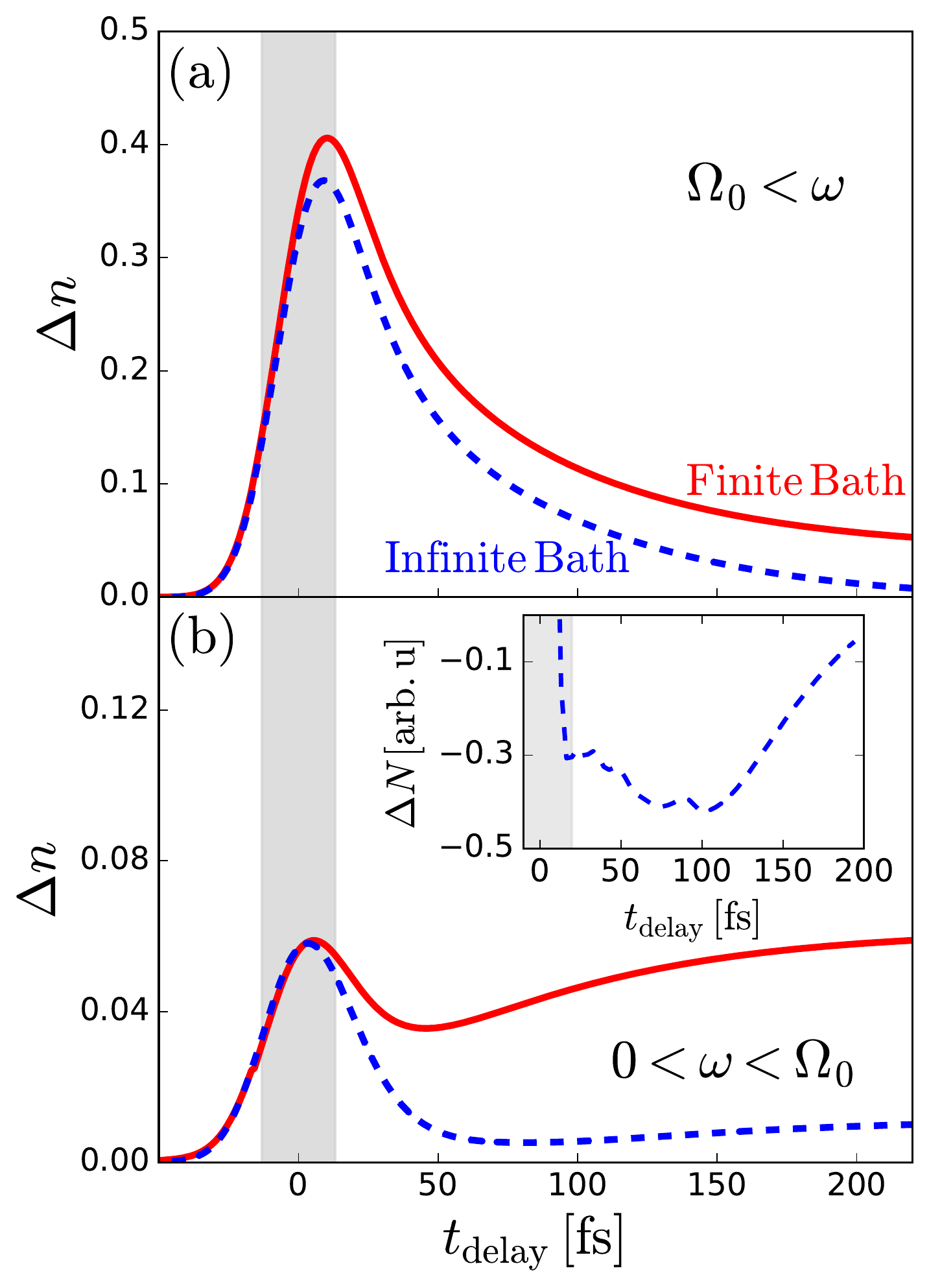}
	\caption{Population dynamics. $\mathrm{(a)}$ Changes in the population above $\Omega_0$ and $\mathrm{(b)}$ inside $\mathcal{W}$ for the Infinite Bath(dashed lines) and the Finite Bath(solid lines) for $\mathrm{F}=1.0$. The inset shows the transient changes of the electronic states inside $\mathcal{W}$. The shaded grey region indicates the time during which the pump is active.}
	\label{fig:pes}
\end{figure}

It was observed in earlier studies\cite{SentefPRX, KemperPRB, YangPRL, KemperTDMD} that the decay of electron population inside the phonon-window defined as $\mathcal{W}=[-\Omega_0,\Omega_0]$ is qualitatively different than the decay outside. The phenomenon was associated with phase space restrictions and was dubbed a ``phonon window effect''.  Here, to study the dynamics of populations inside and outside $\mathcal{W}$, we plot the electron population obtained by summing the spectral weight from tr-ARPES over momenta and energies above $E_\mathrm{F}$ as a function of probe delay time in Fig.~\ref{fig:pes}.

Outside of $\mathcal{W}$, the electrons coupled to the Infinite Bath relax back to the initial state, where the energy injected by the pump has been completely transferred to the Infinite Bath. On the other hand, the electrons coupled to the Finite Bath (solid line) relax toward a new steady state by retaining some of the energy supplied by the pump. 

Inside $\mathcal{W}$, the populations behave more peculiarly. Shortly after the pump field, electrons with energies initially below $\Omega_0$ start to increase their population (Fig.~\ref{fig:pes}(b)). This is true for the Infinite Bath (dashed line) as well as the Finite Bath (solid line) of phonons. 

In the presence of interactions, the electron bare band is modified through the real part of self-energy, and thus so is the density of states (DOS). These modifications are strong primarily around $\pm \Omega_0$ and at the band edges. In equilibrium, we have seen that the DOS is effected by the electron temperature (see Fig.~\ref{fig:dos_equilib}). In nonequilibrium, one can expect that the pump pulse also can effectively alter the density of states. Since the density of states determines the available states to be occupied, we plot the change in the integrated density of states inside $\mathcal{W}$ compared to that in equilibrium as 
\begin{align}
\Delta N(\td) = \int_0^{\Omega_0}d\omega [N(\omega,\td)-N(\omega, -\infty)]
\end{align}
where $N(\omega,\td) = -\frac{1}{\pi}\imgr_\mathrm{loc}(\omega,\td)$. We plot $\Delta N$ for the Infinite Bath in the inset of Fig.~\ref{fig:pes}. After the pump field causes $\Delta N$ to decrease, it slowly restores its equilibrium value resulting in the transiently expanding phase space. Consequently, the number of quasiparticles with $|\omega|<\Omega$ also increases as a function of delay time. Therefore, the transient changes in the density of states are the cause of the population rise in $\mathcal{W}$. In the case of the Finite Bath, where this effect is more pronounced, the situation is more complex. In addition to the transient changes in the DOS, the distribution functions at energies lesser than $\Omega_0$ are found to rise (not shown) via the filling from the states below the Fermi level due to the increased phonon absorption. Therefore, we expect that the combination of these two changes will cause the population rise as is observed in the case of the Finite Bath. This effect might be observed in the experimental studies of population relaxation as a slowing down of relaxation rates for $|\omega|<\Omega$.

\begin{figure}
	\includegraphics[width=0.80\columnwidth]{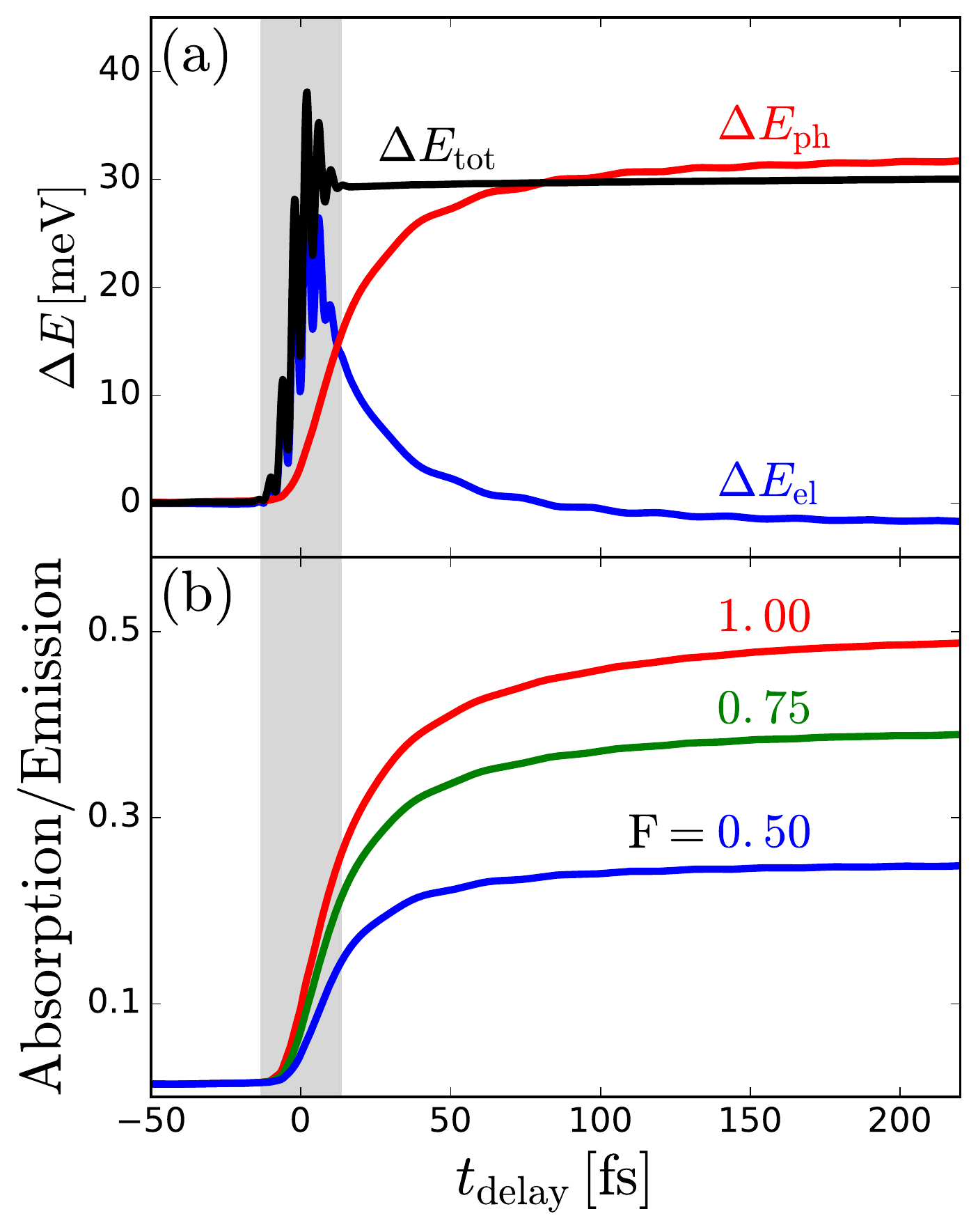}
	\caption{Energy dynamics for the finite bath of phonons. $\mathrm{(a)}$ Temporal changes of the electron energy $\Delta E_\mathrm{el}$ and the phonon energy $\Delta E_\mathrm{ph}$ after the pump deposits $\Delta E_\mathrm{tot}=\Delta E_\mathrm{el}+E_\mathrm{ph}$ energy into the system at $\mathrm{F}=0.5$. $\mathrm{(b)}$ The amount of upscattering (the ratio between absorption and emission) produced during the population relaxation is plotted in panel (b) for different pump fluences.}
	\label{fig:energy}
\end{figure}

When electrons are coupled to the Finite Bath, a bidirectional energy flow between electrons and phonons is enabled. Figure~\ref{fig:energy}(a) shows the energy dynamics of the coupled electron-phonon system.
Initially, the electrons and phonons are in thermal equilibrium before being driven out of equilibrium by the pump field. Then the pump imparts some amount of energy to electrons. Once the pump field shuts off, the electron-phonon system evolves due to a mutual transfer of energy. In the figure, we see the changes in electron energy $\Delta E_\mathrm{el}$, the phonon energy $\Delta E_\mathrm{ph}$, and the total energy of the system $\Delta E_\mathrm{tot}$ as functions of the probe delay time. 

Just after the pump, electrons rapidly relax by emitting phonons causing the phonon number to increase. 
An enhanced phonon occupation stimulates electrons to absorb more phonons and scatter upwards in energy 
which may slow the relaxation process down. In nonequilibrium, we can obtain the phonon number in terms of the lesser component of the phonon Green's function $D^<$:
\begin{align}
\nD(t,t) = \frac{i}{2}\left(D^<(t,t')+\frac{1}{\Omega_0^2}\frac{\partial^2 D^<(t,t')}{\partial t \partial t'}\right)\bigg \vert_{t=t'}-\frac{1}{2}, 
\end{align}
were $\nD$ is the nonequilibrium phonon number, which is equal to the Bose number $\nB$ in equilibrium. 
 The amount of upscattering produced during the relaxation process is given by the ratio $\nD/(\nD+1)$ and is plotted in Fig.~\ref{fig:energy}(b) for various pump fluences. We can see that a substantial amount of upscattering occurs and this will affect the population decay rates.    

\subsubsection{Sum Rules}

The sum rules established for the retarded objects in the Holstein model provides guidance in the analysis of nonequilibrium results.\cite{SumRules} Here, we demonstrate how these rules apply to the nonequilibrium electron self-energy.
  
\begin{figure}
        \includegraphics[width=0.95\columnwidth]{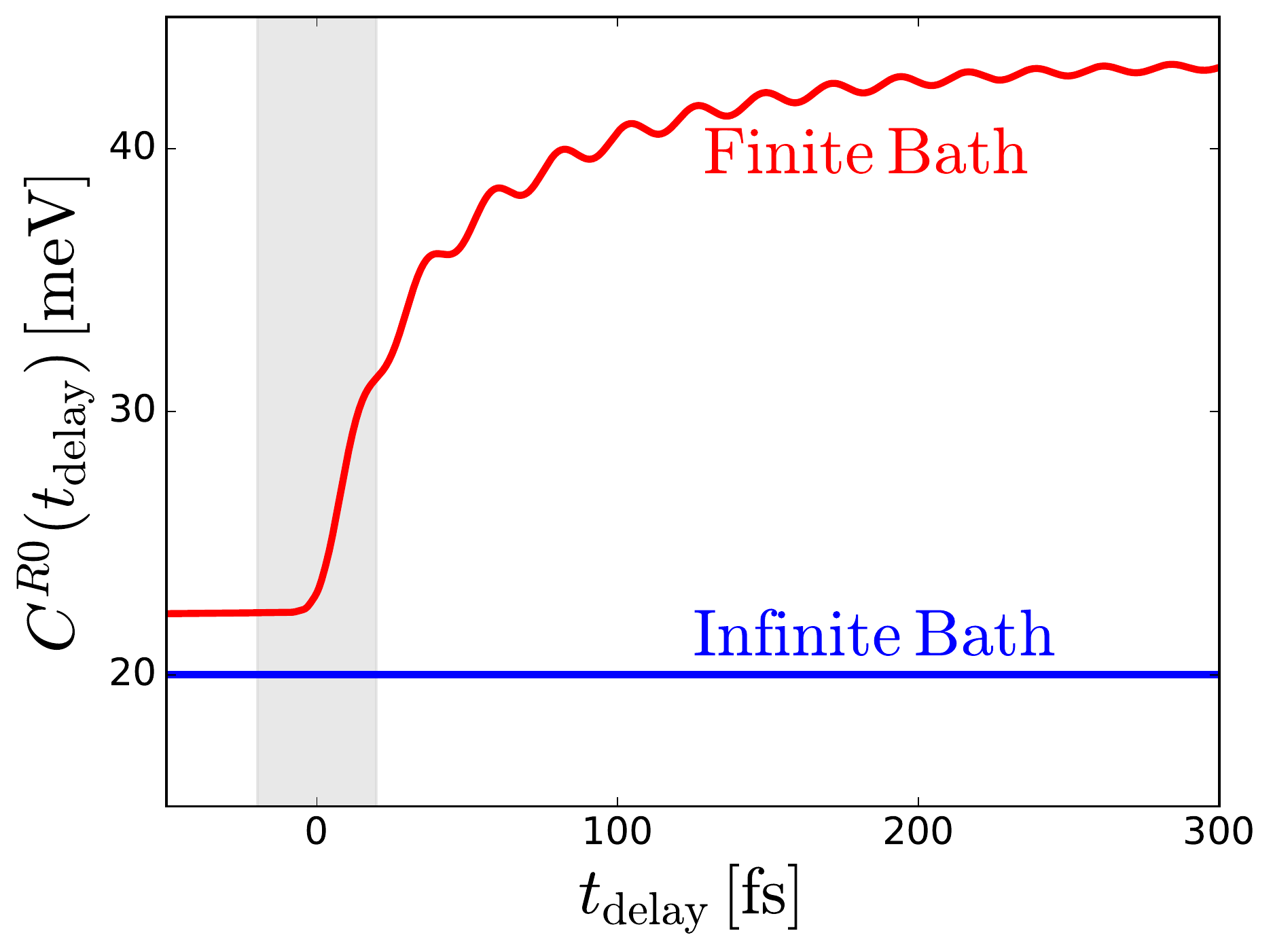}
        \caption{Nonequilibrium sum rule for the zeros moment of the electron self-energy}
        \label{fig:sum_rules}
\end{figure}

In Fig.~\ref{fig:sum_rules}, we plot the zeroth moment of $\imsigma$. The sum rule for the electron self-energy is given by\cite{SumRules}

\begin{align}
C^{R0}(\tave) = -\frac{1}{\pi} &\int d\omega \imsigma (\tave, \omega) \nonumber \\ 
=&g^2[\langle x(\tave)^2\rangle-\langle x(\tave)\rangle ^2],
\end{align}
where $x(t)$ is the phonon displacement operator. This quantity has been suggested as the measure of the electron-phonon interactions in and out of equilibrium and has been shown to be independent of time for the phonons with fixed properties, where it is equal to\cite{KemperPRB}

\begin{align}
C^{R0}(\tave) = g^2[2n_\mathrm{B}(\Omega/T)+1].
\label{eq:first_sum_inf}
\end{align}

Using this formula for the parameters used in our calculation we obtain $20\meV$ confirmed by the numerical result in Fig.~\ref{fig:sum_rules}. On the other hand, for the Finite Bath, one can not obtain the sum rule analytically because the self-energy is computed from the fully dressed Green's functions for electrons and phonons. Nonetheless, in general, we expect that $C^{R0}$, which is proportional to the phonon field fluctuations, is modified due to the feedback of interactions on the phonon bath. This can be seen even more clearly in Fig~\ref{fig:imsigma_equilib} where the area under the curve increases as the temperature of the system is increased and in Fig~\ref{fig:sum_rules}(a) before the pump pulse is active. After the pump pulse, this quantity becomes time-dependent. The magnitude of the phonon field fluctuations increase as a function of delay time (corresponding to the increase in the phonon temperature and consequently the phonon number) and exhibit oscillations with $2\Omega$ frequency.

\subsubsection{Decay Rates}\label{subsec:decay_rates}

\begin{figure*}
	\includegraphics[width=0.70\textwidth]{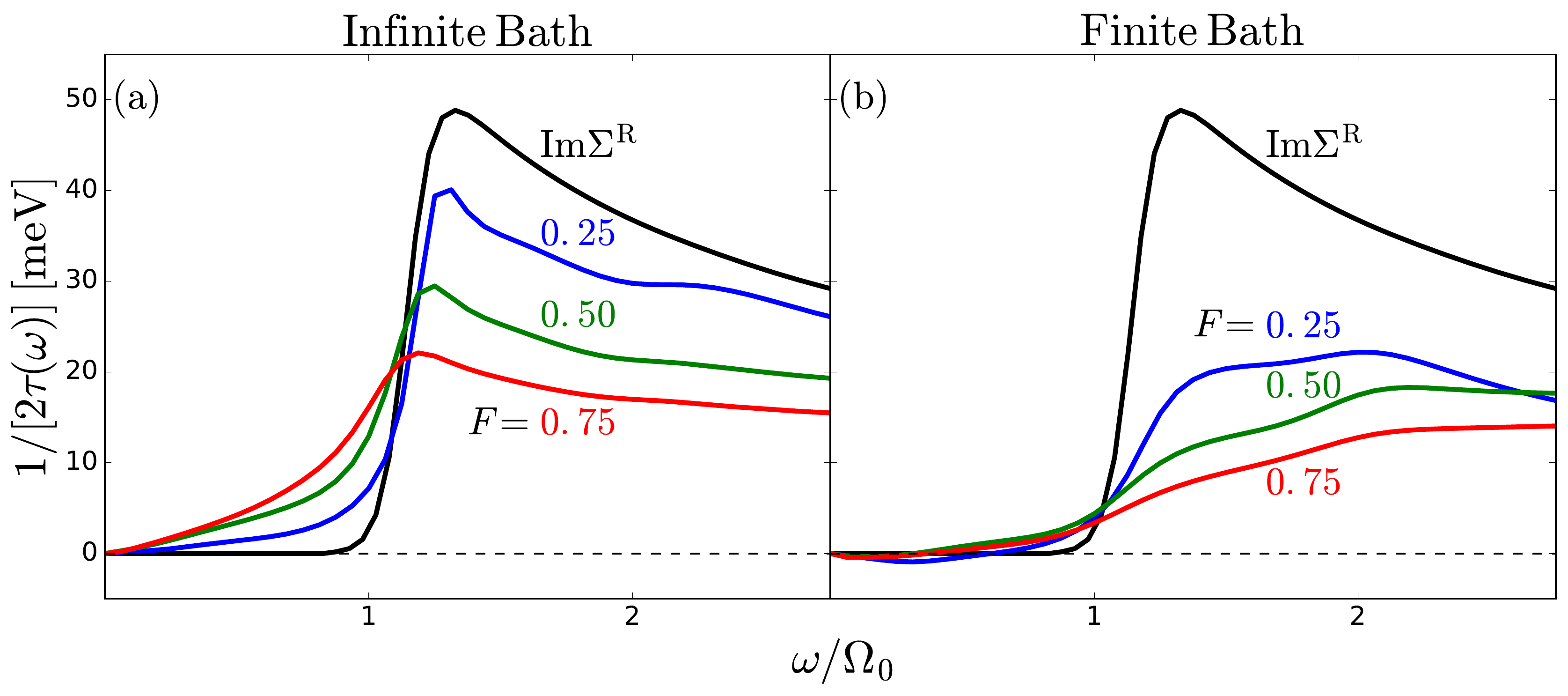}
	\caption{Instantaneous population decay rates as functions of pump fluence for (a) the Infinite Bath and (b) the Finite Bath at a $40\,\mathrm{fs}$ probe delay time. The equilibrium scattering rates are given by $\imsigma$.}
	\label{fig:fluencerates_Omp2}
\end{figure*}

In this section, we study the decay rates of the relaxing electron populations. Before doing so, we note that for the excited population to relax, the system should be connected to a dissipative bath which can draw energy away from the system. In one case, the Infinite Bath assumes this role by absorbing all the excess energy from the nonequilibrium electrons. In the other case, the energy infused by the pump is dynamically exchanged between the electrons and the Finite Bath until some steady state is reached. Comparison of the electron relaxation rates in two cases elucidates the role played by the phonons with dynamically modified properties.

Usually, the decay rates are extracted by fitting the time-dependent population curves to single or multiple exponentials resulting in a constant decay rate. This method ignores the time-dependence of the decay rates and causes possibility of ambiguity in the definition of the decay rates. 
To avoid this, we extract instantaneous decay rates by taking the logarithmic derivative of the momentum integrated population curves at each probe delay time for the reasons outlined in Appendix~\ref{sec:rate_extraction}.
We extract the decay rates for various pump fluences to study the excitation density dependence of the relaxation dynamics. The procedure starts approximately $4\sigma_{\mathrm{p}}$ away from the pump center to avoid the direct influence of pump pulse.  

In Fig.~\ref{fig:fluencerates_Omp2}, we compare the population decay rates of the Finite Bath(right panel) to those of the Infinite Bath(left panel). As shown previously, the pump field can alter the decay rate by modifying the available phase space in the case of the Infinite Bath.\cite{KemperPRB} However, for the Finite Bath, additionally, an enhanced phonon population(see Fig.~\ref{fig:energy}(b)) affects the decay rates through an increased absorption rate. Both the modification in the phase space as well as in the phonon occupation depend on the excitation density. To illustrate this, we have obtained the decay rates for various pump fluences. The rates are extracted at the probe delay time of approximately $40\,\mathrm{fs}$. 

The single-particle scattering rates are obtained from the retarded electron self-energy in equilibrium through $\tau_\mathrm{eq}^{-1}(\omega) = -2\imsigma_\mathrm{eq}(\omega)$. 
Although its equality to the nonequilibrium decay rates is often assumed, only some limiting cases have been 
shown to hold\cite{SentefPRX}, and its validity has been challenged experimentally.\cite{YangPRL, GierzFaraday} Though the presence of other types of interactions (e.g., impurity or Coulomb) renders the equality incorrect, the fundamental difference stems from the fact that the self-energy and the populations decay along different time directions: relative and average time.\cite{KemperAssump} In other words, because the time translational symmetry is broken out of equilibrium, electron populations acquire an average time dependence. On the other hand, the self-energy (encoding the many-body interactions) decay along the relative time $\trel$, which is related to the quasiparticle energy $\omega$ via a Fourier transform.  

\begin{figure*}
	\includegraphics[width=0.70\textwidth]{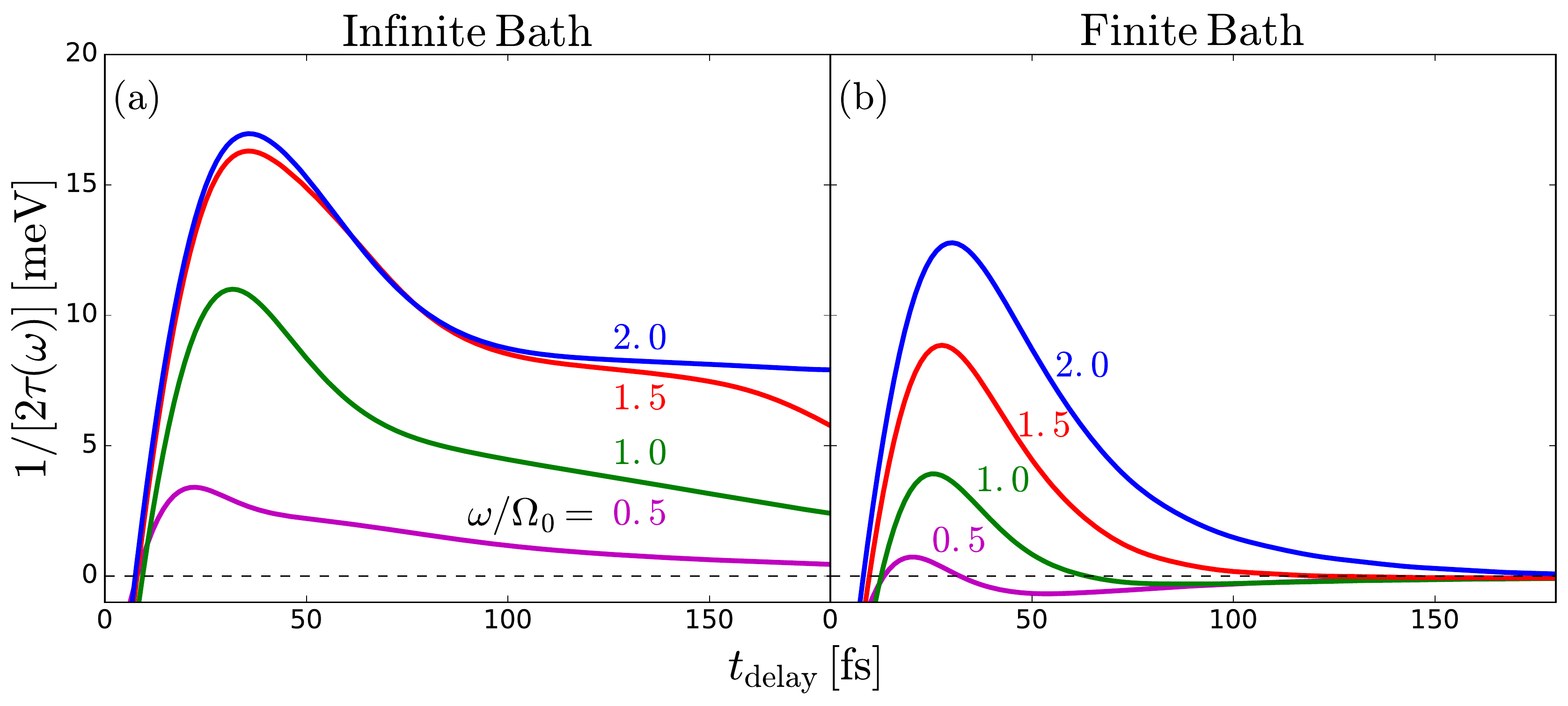}
	\caption{Instantaneous decay rates as functions of probe delay time for (a) the Infinite Bath and (b) the Finite Bath in the vicinity of bare phonon frequency of $\Omega_0$. The pump fluence is $\mathrm{F}=0.5$.}
	\label{fig:timerates}
\end{figure*}

Figure~\ref{fig:fluencerates_Omp2}(a) shows that the rates are fluence-dependent with contrasting trends between the energies inside and outside $\mathcal{W}$ -- the rates increase inside $\mathcal{W}$ and decrease outside with increasing the pump fluence $\FF$. This behavior confirms that the pump induced modifications of the available phase space alter the decay rates.\cite{KemperPRB} At low excitation densities, the population decay rates are described well by the equilibrium scattering rates (or $\imsigma$) for the Infinite Bath. As the excitation density is increased, they start to deviate from the equilibrium rates. On the other hand, for the Finite Bath even at the low fluence regime, the equilibrium rates overestimate the nonequilibrium decay rates (right panel). 

From the comparison of the left and right panels of Fig.~\ref{fig:fluencerates_Omp2}, 
we observe that instantaneous decay rates are suppressed for the Finite Bath relative to those of the Infinite Bath 
at all energies. This indicates that the excited phonon populations cause the decay rates to decrease. 
This is contrary to what happens to the single-particle decay rates i.e. $\imsigma$ in equilibrium (see Fig.~\ref{fig:imsigma_equilib}) where the enhanced phonon population causes them to increase. 
We also observe negative instantaneous decay rates at lower energies which come from the population rise observed in Fig.~\ref{fig:pes}(b). (Note that we are obtaining the rates from a logarithmic derivative as detailed in Appendix~\ref{sec:rate_extraction}.) 
 
\begin{figure*}
	\includegraphics[width=0.70\textwidth]{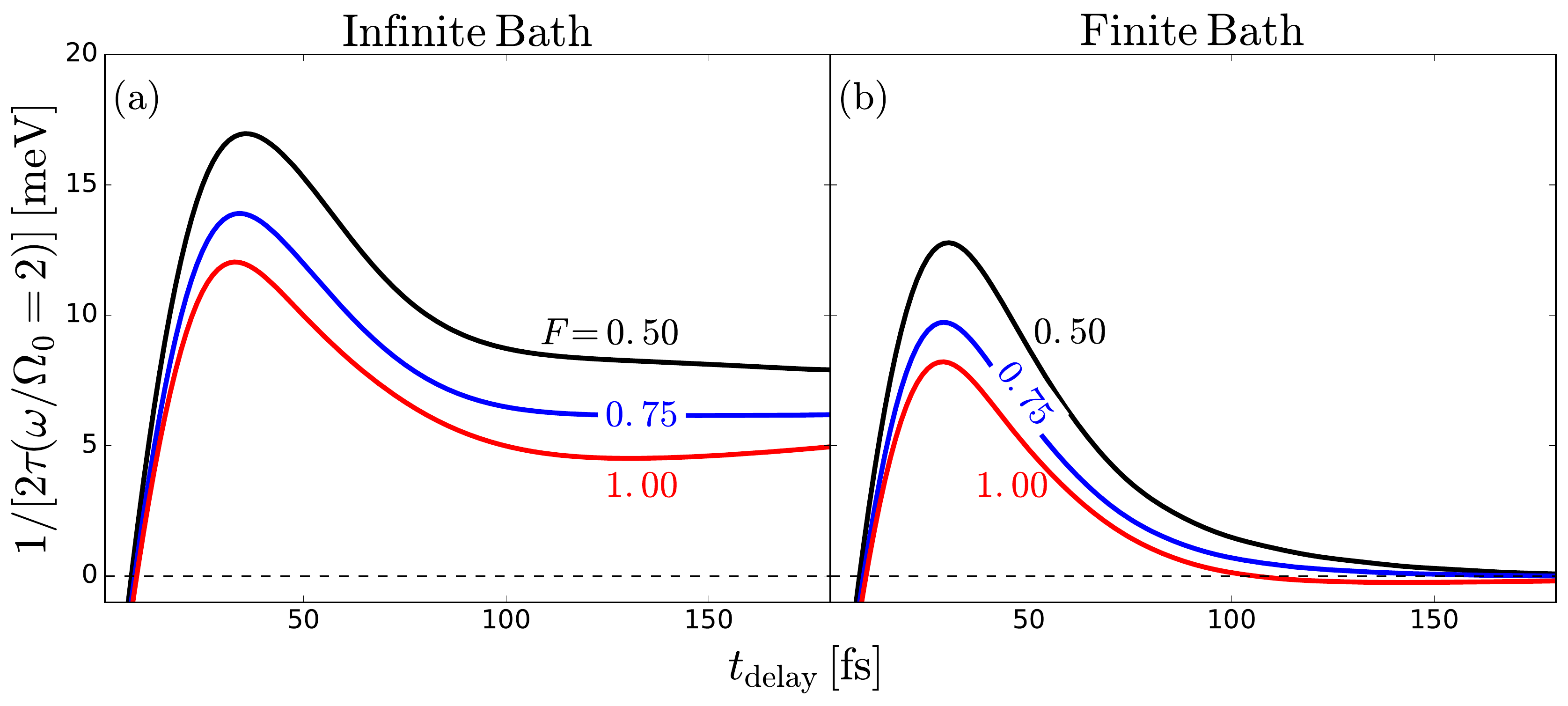}
	\caption{Pump fluence dependence of the decay rates as functions of probe delay time for (a) the Infinite Bath and (b) the Finite Bath at twice the bare phonon frequency. Right: stronger excitation causes the instantaneous decay rates to  decrease faster.}
	\label{fig:timerates_fluence}
\end{figure*}

\begin{figure}
	\includegraphics[width=0.99\columnwidth]{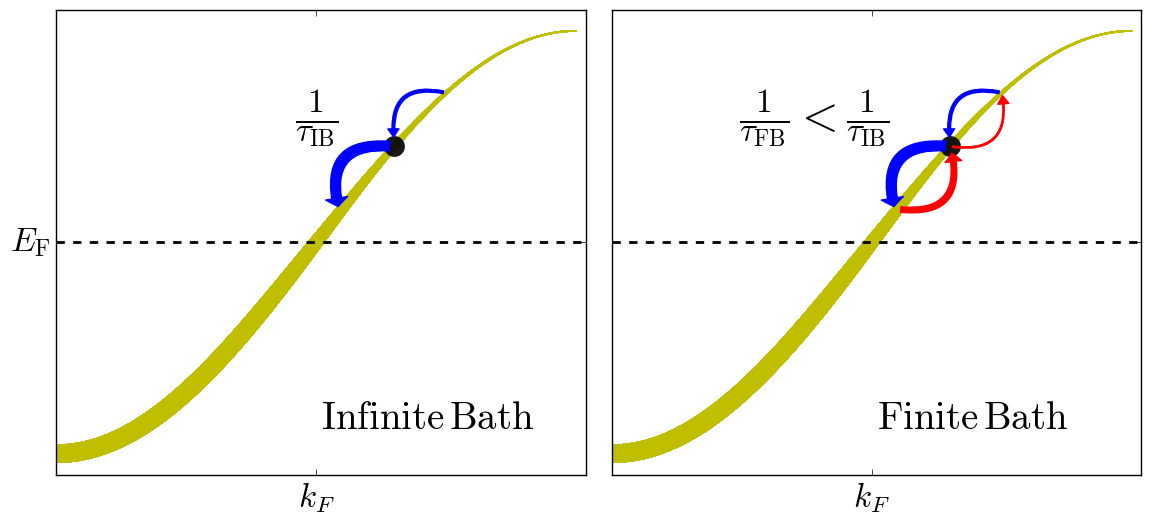}
	\caption{Schematic of how the excited phonons affect the relaxation rates. The width of the yellow curve indicates the density of the excited electrons at a given energy in the band. When the amount of phonons is fixed and small (Infinite Bath), the excited electrons relax mostly via emission of phonons (blue arrows). Allowing the phonons to be excited during the relaxation causes absorption of phonons (red arrows) to increase forcing the electrons to upscatter in energy. These processes, in turn, partially undo the relaxation process itself by suppressing the decay rates.}
	\label{fig:cartoon}
\end{figure}

\begin{figure*}
	\includegraphics[width=0.85\textwidth]{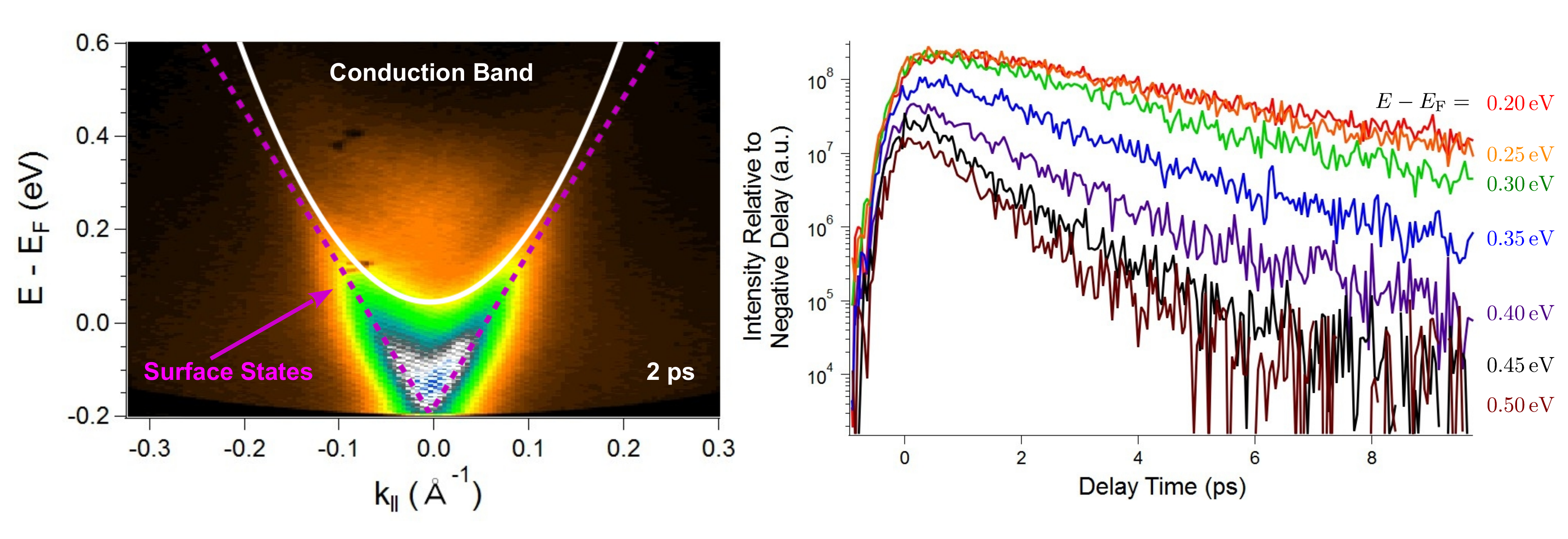}
	\caption{Excited state dynamics of $\mathrm{Bi}_{1.5}\mathrm{Sb}_{0.5} \mathrm{Te}_{1.7}\mathrm{Se}_{1.3}$: (a) tr-ARPES of the topological surface state and the conduction band at $2\,\mathrm{ps}$ probe delay time and (b) the semilog plot of the spectral intensities for the select excitation energies at the topological surface states are shown. The solid and the dashed lines are guides to the eye for the bare surface state and the bare conduction band.}
	\label{fig:experimental_results}
\end{figure*}

As the instantaneous decay rates are affected by the phonon populations which increases as the function of delay time, one can expect the rates will be time-dependent. This is seen in Fig.~\ref{fig:timerates} where we show the rates as a function of time for the energies around $\Omega_0$ for both the Infinite as well as the Finite Bath. This behavior of the decay rates is in stark contrast to the traditional way of obtaining them where they are assigned a constant number through an exponential fit. We observe from the figures that they decrease as functions of $\td$ and approach zero as the final state is reached. Furthermore, the decay rates \textit{decrease} much faster for the Finite Bath due to the transiently enhanced phonon absorption rate. Because the amount of the phonons produced during the relaxation process depends on the energy supplied by the pump (Fig.~\ref{fig:energy}(b)), we also show the time-dependent instantaneous decay rates at different pump fluences in Fig.~\ref{fig:timerates_fluence}. While the decrease in the rates for the Infinite Bath is attributed to the phase space restrictions, the  additional decrease in the rates for the Finite Bath is due to the enhanced absorption. %Thus, the increased pump fluence makes the instantaneous rates to decrease fast and to be strongly time-dependent.

\begin{figure}
	\includegraphics[width=0.99\columnwidth]{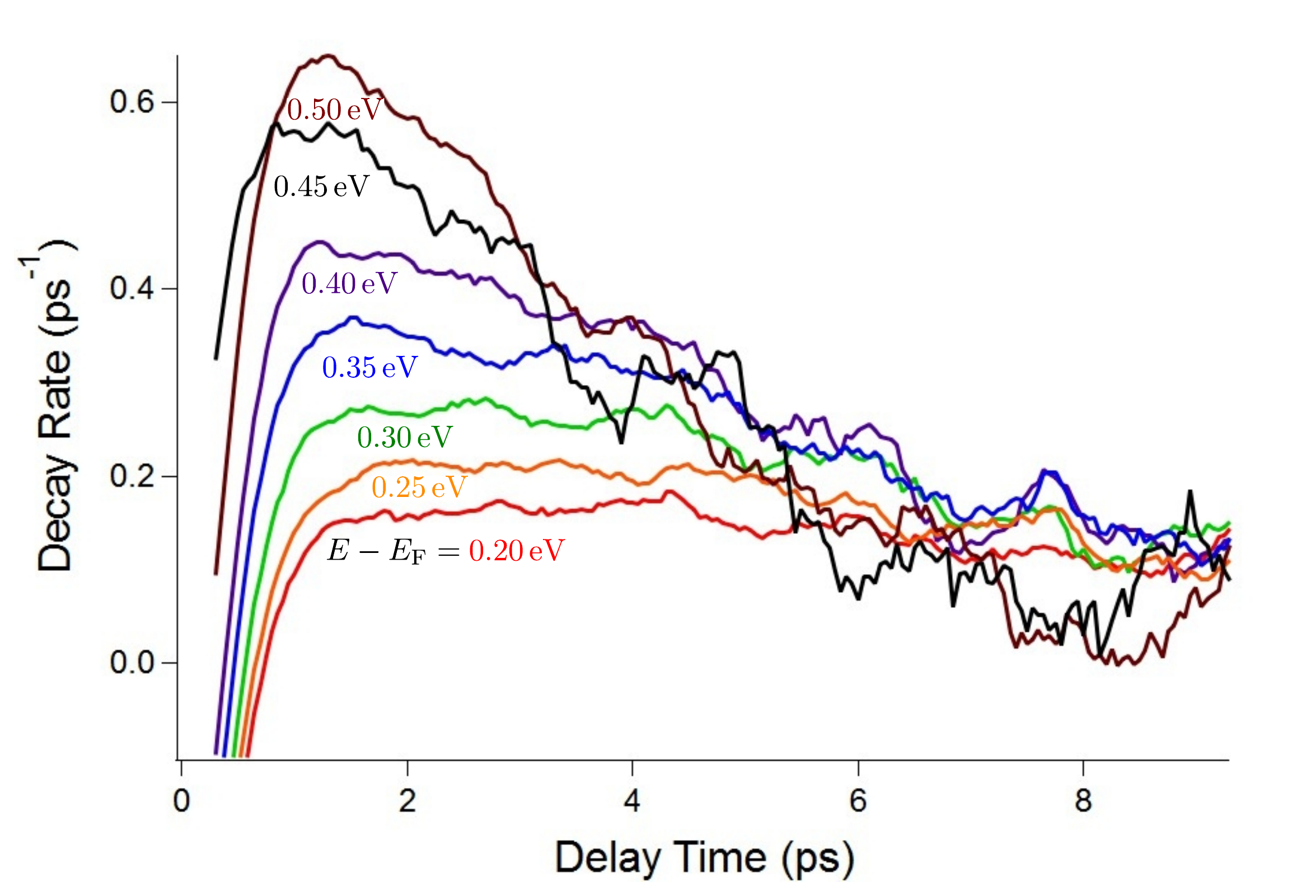}
	\caption{Measured instantaneous decay rates of the excited surface state of $\mathrm{Bi}_2\mathrm{Se}_3$ as a function of probe delay time.}
	\label{fig:experimental_rates}
\end{figure}

The trends observed in the decay rates due to the excited phonon populations can be explained using the schematic in Fig.~\ref{fig:cartoon}. For the Infinite Bath (fixed low phonon temperatures) decay of the excited electrons is enabled mostly by the emptying and filling processes via phonon emission at a given energy above the Fermi level. For the Finite Bath because the number of phonons increases as a function of the delay time, the phonon absorption processes become significant and partially reverse the relaxation itself leading to a slowed decay.    
\subsubsection{Measured Time-Dependent Decay Rates}

Although we presented the time-dependence of the decay rates for the excited electrons in a 2D tight-binding band, the implication of our results is not limited to the specifics of the chosen system. For example, a class of topological insulators offer an illustrative platform to study the relaxation dynamics of excited states in their bulk bands as well as in surface states.\cite{Sobota, SobotaPRL, Glinka} Here, we present experimental results on how relaxation of the excited states of $\mathrm{Bi}_{1.5}\mathrm{Sb}_{0.5} \mathrm{Te}_{1.7}\mathrm{Se}_{1.3}$ evolve as a function of delay time. Because the scale of the decay rates is set by the details of the phonon spectra coupled to the electrons, here, we are just interested in the qualitative dynamics of the excited state relaxation in this particular system.

In order to experimentally observe the excited state dynamics, we performed time- and angle-resolved photoemission spectroscopy (tr-ARPES) on the doped topological insulator $\mathrm{Bi}_{1.5}\mathrm{Sb}_{0.5} \mathrm{Te}_{1.7}\mathrm{Se}_{1.3}$. Our measurements were taken using the output of a regenerative amplifier operating at $250$ kHz and $800$ nm ($1.55$ eV). The system was first pumped with the fundamental pulse and then probed with its fourth harmonic. The time resolution was determined from the full width half max of the rising edge at high energies ($>0.5$ eV above the Fermi level) and is about $600$ fs. Our measurements were performed at room temperature with a Specs Phoibos $150$ analyzer and 2D-CCD detector under a base pressure of $4\times10^{-9}$ mbar.
In order to extract the decay rates as a function of time, we monitored the log of the intensity as a function of delay time obtained from the tr-ARPES of the excited states. ( Figure~\ref{fig:experimental_results}(a) shows the $2\,\mathrm{ps}$ delay time snapshot of the electronic spectra). Examples at various energies are shown in Fig.~\ref{fig:experimental_results}(b). The rate at a particular delay time was then determined from the slope of a linear fit $1.5$ ps long. Plotting the momentum-averaged decay rates as a function of delay time (Fig.~\ref{fig:experimental_rates}), we observe the rates are dynamic, i.e., are time-dependent. The decay rate at a specific energy reaches its peak shortly after the excitation and falls off toward zero as a function of the probe delay time. Qualitatively, this behavior of the decay rates is in agreement with our time-dependent calculations.

\subsubsection{Transient modifications of phonon properties}
	\label{subsec:phonons}

\begin{figure*}
	\includegraphics[width=0.95\textwidth]{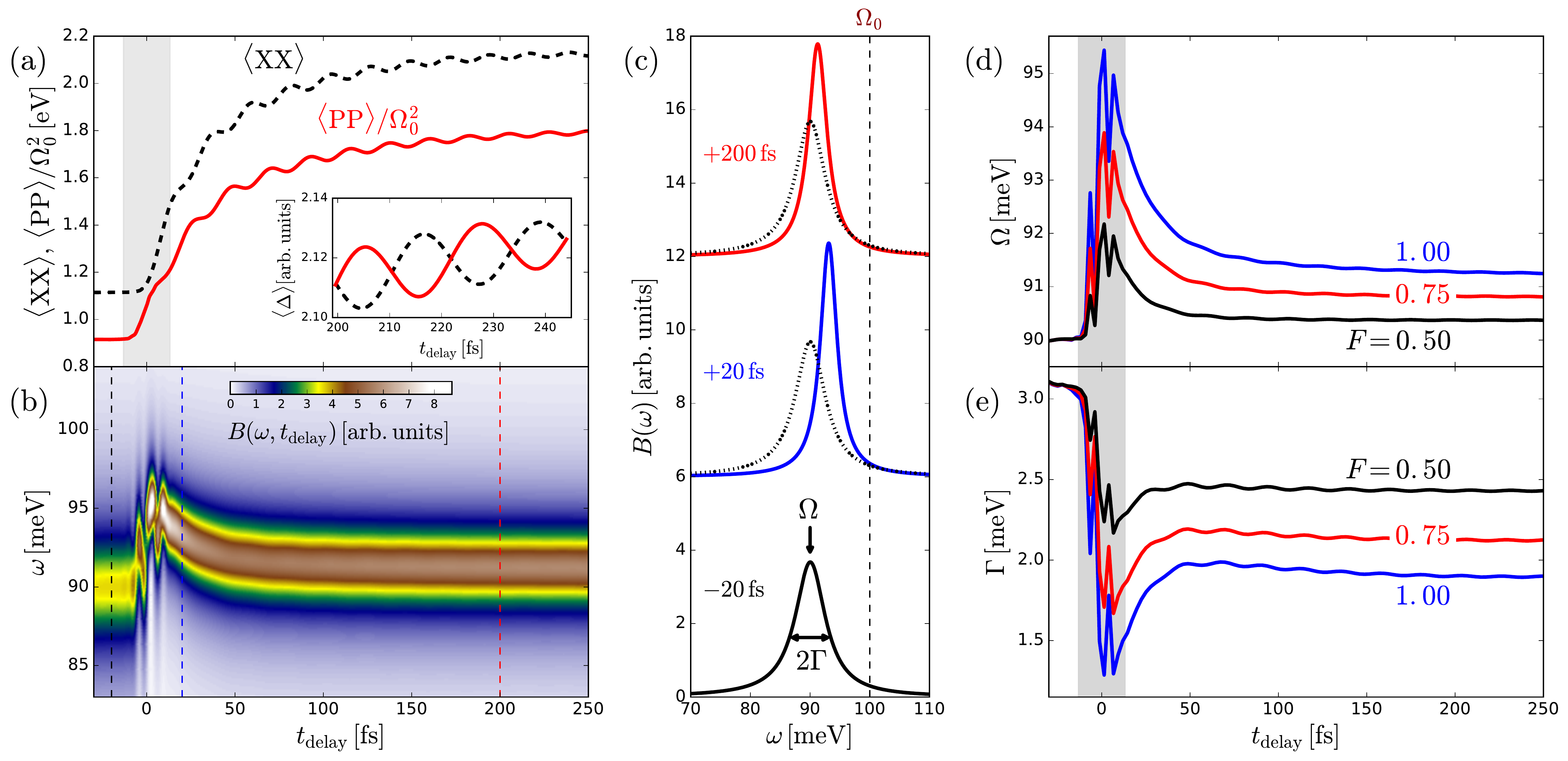}
	\caption{Temporal evolution of phononic properties. $\mathrm{(a)}$ Variances of the lattice displacement $X$ and momentum $P$ as functions of time delay. The inset is a magnified plot showing out-of-phase oscillations in $\langle XX\rangle$ and $\langle PP \rangle$ at the $2\Omega$ frequency. $\mathrm{(b)}$ Time-dependent phonon spectral function $B(\omega,t_\mathrm{delay})$. $\mathrm{(c)}$ Phonon spectral function at marked time slices in $\mathrm{(b)}$: equilibrium, just after, and long after the pump. $\mathrm{(d)}$ Renormalized phonon frequency and $\mathrm{(e)}$ linewidth are extracted from Lorentzian fits to the phonon spectral function at different pump fluences.}
	\label{fig:phonon}
\end{figure*}

In the electronic spectral function, we have seen the signatures of the phononic properties through the size and the shape of the kink. In our study, the phonon properties are affected indirectly by the pump field. In the case of the Finite Bath, the Einstein phonons acquire a finite width and a renormalized frequency because of the interaction with the continuous electronic states. This is reflected in the fact that the position of the kink was shifted upward indicating the softening of the phonon frequency. 

In Fig.~\ref{fig:phonon}(a), we observe the transient changes in the atomic mean square displacements $\langle XX\rangle$ and its conjugate variable $\langle PP\rangle$. The shaded region demarcates the times when the pump field is on. In addition to the increase in their values, both oscillate at the twice the phonon frequency but with a $\pi$ phase shift. The inset shows these oscillations on a larger scale. These oscillations at $2\Omega$ frequency have also been observed in interaction quench study of the Holstein model.\cite{Murakami}

The phonon spectral function is given by the imaginary part of the phonon Green's function
\begin{align}
B(\omega,\td) = -\frac{1}{\pi}\imdr(\omega,\td). 
\end{align}
For the coupling strength used, $D^\mathrm{R}$ decays on much longer timescales that are not numerically feasible to reach. Thus, we obtain the phonon spectral function via the retarded phonon self-energy $\Pi^\mathrm{R}$ (see Eq.~\ref{eq:phonon_self}) after a Wigner rotation and Fourier transform along $\trel$. In Fig.~\ref{fig:phonon}(b), we plot the evolution of the spectral function before and after the pump pulse. To compare the modification in its profile, we compare time snapshots in equilibrium and after the excitation (Fig.~\ref{fig:phonon}(c)). 
The vertical dashed line indicates the value of the bare phonon frequency $\Omega_0$ without interactions 
taken into account. In equilibrium($\td =-20 \fs$), the phonon frequency is softened to $\Omega$ and 
acquires a finite lifetime $\Gamma$ because of the interactions. After the optical excitation ($\td = 20 \fs$), 
the phonon frequency and the lifetime are renormalized transiently which evolve toward the long time steady 
values($\td = 200 \fs$). This shows that the pump field can effectively weaken the changes caused by 
the strong interactions between electrons and phonons. Fig.~\ref{fig:phonon}(d/e) displays the pump induced 
changes in frequency/linewidth extracted from the spectral function at different pump fluences. Long after the excitation, these quantities evolve toward modified values different from the pre-pump values. These modifications are accentuated when the pump fluence is increased. Recent experimental studies have demonstrated the importance of the pump-induced changes of phonon properties in understanding the nature of charge stripes in complex transition oxides.\cite{Coslovich}   

%%%%%%%%%%%%%%%%%%%%%%%%%%%%%%%%%%%%%%%%%%%%%%%%%%%%

\section{Conclusions}
\label{sec:conclusion}

In this work, we have studied the effect of exciting phonon populations on the relaxation dynamics of nonequilibriuim electrons after an ultrafast optical excitation. The time-dependent suppression of the relaxation rates caused by the excited phonons is the central result of the paper.

We performed tr-ARPES of the nonequilibrium electrons to investigate their spectral dynamics. Solving the Dyson equation for the phonon properties enabled us to account for excited phonon populations and to explain how they affect the relaxation dynamics compared to the case where the phonon properties are fixed. In equilibrium, we do not find significant changes in the spectra when the phonon properties are accounted for self-consistently. In contrast, in nonequilibrium, electrons exhibit starkly different single-particle as well as the population dynamics when the phonons are excited. Particularly, the nonequilibrium electrons interacting with the self-consistently modified phonons reach a new final state different from the pre-pumped equilibrium state. Although the final state can be obtained using equilibrium ARPES at an elevated temperature, the transient spectra and their dynamics are nonequilibrium in nature, so they must be treated accordingly. The extracted energy-dependent decay rates displayed strong suppression when the phonons are excited. This is because the phonons produced during the relaxation process cause the absorption rates to go up and partially undo the relaxation resulting in the suppressed decay rates. We demonstrate that this effect becomes more prominent with the increase of the pump fluence where more energy is pumped into the system causing more phonons to be produced during the relaxation. Because the phonon population increases as a function of the probe delay time, the rates become strongly time-dependent. To support this, we measured the excited state decay rates in the topological surface state of $\mathrm{Bi}_{1.5}\mathrm{Sb}_{0.5} \mathrm{Te}_{1.7}\mathrm{Se}_{1.3}$, and we observed time-dependent decay rates in qualitative agreement with our calculations.

The dynamic changes in the electronic properties due to the excited phonon populations concomitantly affects the phononic properties as well. This is observed in the transiently stiffened phonon frequency and narrowed linewidth after the pump pulse. Tailoring the crystal lattice properties by laser pulses widens the possibilities towards materials by design in ultrafast timescales.

The framework used in this work can be also applied to the ordered states of matter such as superconductivity or the charge density waves in nonequilibrium, which are often understood in the light of a phenomenological Rothwarf-Taylor model, to clarify the role of the phonon bottleneck effect in the dynamics of the quasiparticles and the Cooper pairs. These will be subjects of future studies. 

%%%%%%%%%%%%%%%%%%%%%%%%%%%%%%%%%%%%%%%%%%%%%%%%
\begin{acknowledgments}
The experimental work was supported by the U.S. Department of Energy, Office of Science, Basic Energy Sciences under award No. DE-SC0010324 and instrumentation funding was provided by a UNC-GA ROI grant. J.K.F. was supported by the U.S. Department of Energy, Office of Basic Energy Sciences, Materials Sciences and Engineering under Contract No. DE-FG02-08ER46542 and also by the McDevitt bequest at Georgetown. Computational resources were provided by the
National Energy Research Scientific Computing Center supported by the U.S. Department of Energy, Office of
Science, under Contract No. DE-AC02-05CH11231. Helpful discussions with Ankit Kumar is acknowledged.
\end{acknowledgments}
%%%%%%%%%%%%%%%%%%%%%%%%%%%%%%%%%%%%%%%%%%%%%%%%
\appendix

\begin{figure*}[htpb]
\includegraphics[width=0.90\textwidth]{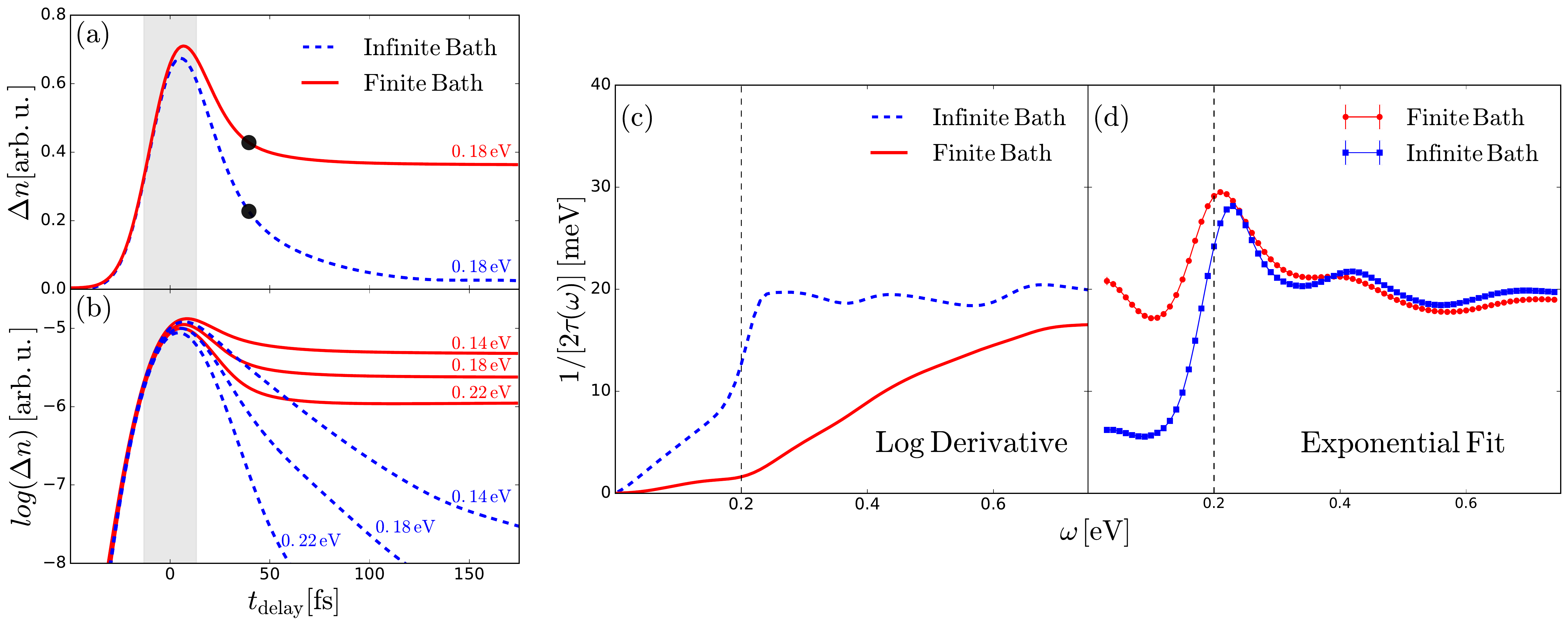}
\caption{Comparison of the decay rates extracted using $\mathrm{(c)}$ logarithmic derivative and $\mathrm{(d)}$ fits to exponentially decaying functions. Transient population curves are given for the energies around the phonon frequency ($\mathrm{a,b}$). The decay rates extracted around the time points are indicated by the black circles.} 
	\label{fig:extractrates}
\end{figure*}

	\section{Exponential Fit vs. Logarithmic Derivative}\label{sec:rate_extraction}
Typical pump-probe curves consist of a rapid rise in some observable, followed by a slower decrease as the system returns to some new steady state.  One example of this is the two-temperature model (TTM) for electrons coupled to phonons, where a rapid increase in the electronic temperature $T_e$ is followed by a steady decay in $T_e$ and a concomitant rise in the phonon temperature $T_p$, which continues until the temperatures are equal. When faced with a such typical pump-probe population relaxation curve, a common analysis method is to resort to exponential fits of the form
\begin{align}
\Delta y(t) = A e^{-\gamma t} + B,
\label{eq:exp}
\end{align}
where $A$ and $B$ are constants representing some measure of the time-dependent part of $\Delta y(t)$ and a final-state offset that represents some difference from the pre-pump state, respectively, and $\gamma$ is the corresponding rate(inverse time constant $\tau^{-1}$) of this change.  The fits are typically good, and some note is made of the dependence of the rate $\gamma$ on experimental parameters.

It is worth taking a step back and considering the theoretical basis for such modeling. The exponential decay arises from a differential equation
\begin{align}
\frac{d \Delta y(t)}{dt} = - \gamma\cdot \Delta y(t).
\label{eq:de}
\end{align}
The solution is indeed an exponentially decaying function, although since it is a first order equation it only has a single constant of integration, nominally $A$. The final-state offset $B$ is introduced because, quite commonly, the final state is \textit{different} from the pre-pump initial state.  However, as we will see, the inclusion of the offset $B$ complicates the analysis of the rates.

Formally, Eq.~\ref{eq:de} cannot achieve a different final state than the initial state \textemdash\ it must decay to $0$ at long times. A different final state arises because some other process is playing a role. Physically, this is quite sensible; for example in the TTM, $T_p$ is rising, which leads to a different final state.  Within the differential equation, this may be modeled through a time-dependent rate:
\begin{align}
\frac{d \Delta y(t)}{dt} = - \gamma(t)\cdot\Delta y(t).
\label{eq:de2}
\end{align}
At this point, we may already conclude that a simple exponential fit does not capture the correct dynamics since the rate depends on time, although as a fit it often works quite well.
However, a more serious problem that arises from exponential fitting with an offset is an ambiguity in the definition of the rate \textemdash\ or to put this another way, which question is being asked.  To get an idea about the underlying mechanism of the decay, one may ask the question: ``How quickly is the population changing at a particular time?''  By performing an exponential fit with a final offset, this question becomes ``How quickly does the population reach its new final state?''  The answers to these questions may be vastly different, in particular if the final offset $B$ is close to the maximum change in $\Delta y(t)$. Which question is the important one depends on the points one wishes to make. Instead, the decay appears to be faster because the final state is simply closer by, and the exponential fit answers the second question, ``How quickly does the population reach its new final state?'' Ideally, we would like to know the population decay rate independent of the final state, since this gives a true measure of the interactions that cause the decay. The first question, ``How quickly is the population changing at a particular time'' can be obtained by reverting to the differential equation, and evaluating
\begin{align}
\gamma(t) = -\frac{1}{\Delta y(t)} \frac{d \Delta y(t)}{dt} = - \frac{d}{dt} \log\left[\Delta y(t)\right].
\end{align}
By construction, this evaluation returns the expected instantaneous decay rate $\gamma(t)$ from Eq.~\ref{eq:de2}. 
Analogously, we obtain the decay rates at different quasiparticle energies $\omega$ via the expression   
\begin{align}
\frac{1}{\tau}(\tave,\omega)= -\frac{\partial}{\partial\tave}\log\left[ \sum_\kk \delta I(\kk,\tave,\omega)\right]. 
\end{align}
Here, $\delta I$ is the change of the tr-ARPES intensity with respect to its equilibrium value. 

We illustrate this point by comparing the extracted decay rates using both methods for the sample transient population data in Fig.~\ref{fig:extractrates}. Two curves plotted in Fig.~\ref{fig:extractrates}(a) represent the time traces of populations coupled to the optical phonons of bare frequency $\Omega_0=0.2\,\mathrm{eV}$ at quasiparticle energy $\omega=0.18\,\mathrm{eV}$ for the Infinite Bath and the Finite Bath. The curves reach very different final states long after the pump field. We extract the decay rates of these populations around the time point shown by the black circles using both methods. In Fig.~\ref{fig:extractrates}(b), we also show these for the energies around the phonon frequency in a semilog scale, where the magnitude of the decay rates corresponds to the slope of a curve. Here, we observe that the rates for the Finite Bath are much less at all probe delay times, and become independent of time soon after the excitation.  

The decay rates obtained for all energies using the logarithmic derivative method mirror the trend observed in Fig.~\ref{fig:extractrates}(b). Here, the decay rates of the populations coupled to the Finite Bath are slower at all energies. On the other hand, contrary to the latter, using the exponential fits, we obtain faster decay rates for the Finite Bath than those for the Infinite Bath at lower energies and similar decay rates at higher energies. In conclusion, depending on the method used to extract the decay rates, one can obtain different results. Because we are interested in addressing the question ``How quickly is the population changing at a particular time at a particular energy``, we use the method of logarithmic derivatives to determine instantaneous decay rates in this work. 

%%%%%%%%%%%%%%%%%%%%%%%%%%%%%%%%%%%%%%%%%%%%%%%%

\bibliography{ref}

\end{document}